\documentclass[fleqn,usenatbib]{mnras}

\usepackage{newtxtext,newtxmath}

\usepackage[T1]{fontenc}

\DeclareRobustCommand{\VAN}[3]{#2}
\let\VANthebibliography\thebibliography
\def\thebibliography{\DeclareRobustCommand{\VAN}[3]{##3}\VANthebibliography}

\usepackage{graphicx}	% Including figure files
\usepackage{amsmath}	% Advanced maths commands
\usepackage{svg}
\let\oldAA\AA
\renewcommand{\AA}{\text{\normalfont\oldAA}}

\usepackage{bm}
\usepackage{url}

\usepackage{lineno}
\usepackage{booktabs}
\usepackage{placeins}
\usepackage{orcidlink}
\usepackage{hyperref}
\usepackage{xcolor}

\title[AT2019cmw]{AT2019cmw: A highly luminous, cooling featureless TDE candidate from the disruption of a high mass star in an early-type galaxy}

\author[J. Wise et al.]{
Jacob L. Wise,\orcidlink{0000-0003-0733-2916}$^{1}$\thanks{E-mail: \url{J.L.Wise@2022.ljmu.ac.uk}}
Daniel A. Perley,\orcidlink{0000-0001-8472-1996}$^{1}$
Nikhil Sarin,\orcidlink{0000-0003-2700-1030}$^{2}$
Tatsuya Matsumoto,\orcidlink{0000-0002-9350-6793}$^{3,4,5}$
K-Ryan Hinds,\orcidlink{0000-0002-0129-806X}$^{1}$
\newauthor
Yuhan Yao,\orcidlink{0000-0001-6747-8509}$^{6,7}$
Jesper Sollerman,\orcidlink{0000-0003-1546-6615}$^{8}$
Steve Schulze,\orcidlink{0000-0001-6797-1889}$^{9}$
Aleksandra Bochenek,\orcidlink{0009-0008-2714-2507}$^{1}$
Michael W. Coughlin,\orcidlink{0000-0002-8262-2924}$^{10}$
\newauthor
Kishalay De,\orcidlink{0000-0002-8989-0542}$^{11}$
Richard Dekany,\orcidlink{0000-0002-5884-7867}$^{12}$
Sara Frederick,\orcidlink{0000-0001-9676-730X}$^{13}$
Christoffer Fremling,\orcidlink{0000-0002-4223-103X}$^{12,14}$
Suvi Gezari,\orcidlink{0000-0003-3703-5154}$^{15,16}$
\newauthor
Matthew J. Graham,\orcidlink{0000-0002-3168-0139}$^{17}$
Anna Y. Q. Ho,\orcidlink{0000-0002-9017-3567}$^{18}$
Shrinivas Kulkarni,\orcidlink{0000-0001-5390-8563}$^{17}$
Russ R. Laher,\orcidlink{0000-0003-2451-5482}$^{19}$
Conor Omand,\orcidlink{0000-0002-9646-8710}$^{1}$
\newauthor
Natalya Johnson,\orcidlink{0009-0008-8062-445X}$^{20}$
Yashvi Sharma,\orcidlink{0000-0003-4531-1745}$^{17}$
Kirsty Taggart,\orcidlink{0000-0002-5748-4558}$^{21}$
Charlotte Ward,\orcidlink{0000-0002-4557-6682}$^{22}$
Avery Wold,\orcidlink{0000-0002-9998-6732}$^{19}$
Lin Yan\orcidlink{0000-0003-1710-9339}$^{12}$
\\
$^{1}$Astrophysics Research Institute, Liverpool John Moores University, 146 Brownlow Hill, Liverpool L3 5RF, UK\\
$^{2}$Oskar Klein Centre for Cosmoparticle Physics, Department of Physics, Stockholm University, AlbaNova, Stockholm SE-106 91, Sweden\\
$^{3}$Department of Astronomy, Kyoto University, Kitashirakawa-Oiwake-cho, Sakyo-ku, Kyoto, 606-8502, Japan\\
$^{4}$Hakubi Center, Kyoto University, Yoshida-honmachi, Sakyo-ku, Kyoto, 606-8501, Japan\\
$^{5}$Department of Astronomy, School of Science, the University of Tokyo, Bunkyo-ku, Tokyo 113-0033, Japan\\
$^{6}$Miller Institute for Basic Research in Science, 206B Stanley Hall, Berkeley, CA 94720, USA\\
$^{7}$Department of Astronomy, University of California, Berkeley, CA 94720-3411, USA\\
$^{8}$Oskar Klein Centre, Department of Astronomy, Stockholm University, AlbaNova, SE-106 91 Stockholm, Sweden\\
$^{9}$Center for Interdisciplinary Exploration and Research in Astrophysics (CIERA), Northwestern University, 1800 Sherman Road, Evanston, IL60201, USA\\
$^{10}$School of Physics and Astronomy, University of Minnesota, Minneapolis, Minnesota 55455, USA\\
$^{11}$MIT Kavli Institute for Astrophysics and Space Research, 70 Vassar St., Cambridge, MA 02139, USA\\
$^{12}$Caltech Optical Observatories, California Institute of Technology, Pasadena, CA 91125, USA\\
$^{13}$Vanderbilt Department of Physics and Astronomy, 1221 Stevenson Center Lane, Nashville, TN, 37240\\
$^{14}$Division of Physics, Mathematics and Astronomy, California Institute of Technology, Pasadena, CA 91125, USA\\
$^{15}$Space Telescope Science Institute, 3700 San Martin Dr, Baltimore, MD 21218, USA\\
$^{16}$Physics and Astronomy Department, Johns Hopkins University, Baltimore, MD 21218, USA\\
$^{17}$Cahill Center for Astrophysics, California Institute of Technology, MC 249-17, 1200 E California Boulevard, Pasadena, CA 91125, USA\\
$^{18}$Department of Astronomy, Cornell University, Ithaca, NY 14853, USA\\
$^{19}$IPAC, California Institute of Technology, 1200 E. California Blvd, Pasadena, CA 91125, USA\\
$^{20}$Department of Physics, Drexel University, Philadelphia, PA 19104, USA\\
$^{21}$Department of Astronomy and Astrophysics, University of California, Santa Cruz, CA 95064, USA\\
$^{22}$Department of Astrophysical Sciences, Princeton University, Princeton, NJ 08544, USA\\
}
\date{Accepted XXX. Received YYY; in original form ZZZ}

\pubyear{2026}

\begin{document}
\label{firstpage}
\pagerange{\pageref{firstpage}--\pageref{lastpage}}
\maketitle

% Abstract of the paper
\begin{abstract}
We present optical/UV photometric and spectroscopic observations, as well as X-ray and radio follow-up, of the extraordinary event AT2019cmw. With a peak bolometric luminosity of $\mathrm{\sim10^{45.6}\,erg\,s^{-1}}$, it is one of the most luminous thermal transients ever discovered. Extensive spectroscopic follow-up post-peak showed only a featureless continuum throughout its evolution. This, combined with its nuclear location, blue colour at peak and lack of prior evidence of an AGN in its host lead us to interpret this event as a “featureless” tidal disruption event (TDE). It displays photometric evolution atypical of most TDEs, cooling from $\mathrm{\sim30\,kK}$ to $\mathrm{\sim10\,kK}$ in the first $\mathrm{\sim300}$ days post-peak, with potential implications for future photometric selection of candidate TDEs. No X-ray or radio emission is detected, placing constraints on the presence of on-axis jetted emission or a visible inner-accretion disk. Modelling the optical light curve with existing theoretical prescriptions, we find that AT2019cmw may be the result of the disruption of a star in the tens of solar masses by a supermassive black hole (SMBH). Combined with a lack of detectable star formation in its host galaxy, it could imply the existence of a localised region of star formation around the SMBH. This could provide a new window to probe nuclear star formation and the shape of the initial mass function (IMF) in close proximity to SMBHs out to relatively high redshifts.

\end{abstract}

% Select between one and six entries from the list of approved keywords.
% Don't make up new ones.

%%%%%%%%%%%%%%%%%%%%%%%%%%%%%%%%%%%%%%%%%%%%%%%%%%

%%%%%%%%%%%%%%%%% BODY OF PAPER %%%%%%%%%%%%%%%%%%

\section{Introduction}
\label{sec:Introduction}

In recent years, wide-area sky surveys have enabled the discovery of thousands of astronomical transients. Current on sky facilities include the Zwicky Transient Facility (ZTF; \citealt{bellm_unblinking_2017,bellm_zwicky_2019,graham_zwicky_2019,masci_zwicky_2019,dekany_zwicky_2020}), the Gravitational-wave Optical Transient Observer (GOTO; \citealt{steeghs_gravitational-wave_2022}), the Asteroid Terrestrial Last Alert System (ATLAS; \citealt{tonry_atlas_2018,smith_design_2020}), the All-Sky Automated Survey for Supernovae
(ASAS-SN; \citealt{shappee_man_2014,kochanek_all-sky_2017,hart_asas-sn_2023}), the Catalina Sky Survey (CSS; \citealt{larson_catalina_1998,larson_css_2003} and associated Catalina Real-time Transient Survey (CRTS; \citealt{drake_first_2009,djorgovski_catalina_2011,mahabal_discovery_2011}), and Pan-STARRS1 (PS1; \citealt{chambers_pan-starrs1_2016, flewelling_pan-starrs1_2020}). ZTF images the entire northern sky in both $g$ and $r$ filters on a $\sim48$ hour cadence down to an apparent magnitude of $\sim20.5$ \citep{bellm_unblinking_2017,bellm_zwicky_2019}, and includes a dedicated facility for spectroscopic follow-up \citep{blagorodnova_sed_2018,rigault_fully_2019}. This enables complete population-level studies of astrophysical transients, such as the Bright Transient Survey (BTS) - a large, purely magnitude-limited classification survey that has so far accumulated over 10000 classified events. \citep{fremling_zwicky_2020,perley_zwicky_2020,rehemtulla_zwicky_2024}.

In addition to classifying large numbers of supernovae (SNe) belonging to well-established types, ZTF has also discovered and characterized significant numbers of much rarer transients that may require energy sources or explosion mechanisms distinct from normal thermonuclear and core-collapse SNe. Examples include hydrogen-poor Type I Superluminous Supernovae (SLSNe) \citep{gal-yam_most_2019, inserra_statistical_2018} and Luminous Fast Blue Optical Transients (LFBOTs) \citep{drout_rapidly_2014, perley_fast_2019, margutti_embedded_2019, metzger_luminous_2022, ho_search_2023, chrimes_at2023fhn_2024}.

Similarly, an increasing number of Tidal Disruption Events (TDEs), the flares produced when a star passes sufficiently close to a black hole to be torn apart and subsequently accreted \citep{rees_tidal_1988,phinney_manifestations_1989}, have been categorised. Observationally, TDEs have been discovered at a variety of electromagnetic wavelengths, including in the infrared \citep{reynolds_energetic_2022,masterson_new_2024}, UV/optical \citep{van_velzen_first_2019,van_velzen_seventeen_2021,hammerstein_final_2023,yao_tidal_2023}, soft X-ray \citep{sazonov_first_2021,grotova_population_2025,oconnor_characterization_2025}, and hard X-ray \citep{bloom_possible_2011,yao_-axis_2024,ho_luminous_2025}. Candidate TDE neutrino associations have also been proposed \citep{stein_tidal_2021,reusch_candidate_2022,jiang_two_2023,yuan_at2021lwx_2024,vanvelzen_establishing_2024}, suggesting the possibility for future multi-messenger selection of TDEs. Additionally, other recently discovered classes of nuclear transients in galaxies have been proposed to be potentially caused by TDEs, such as "Ambiguous Nuclear Transients" (ANTs; \citealt{trakhtenbrot_new_2019,oates_swiftuvot_2024,wiseman_systematically_2025}) and "Extreme Nuclear Transients" (ENTs; \citealt{hinkle_most_2025}), which appear spectrally similar to Active Galactic Nuclei (AGN).

Observationally, optically-selected TDEs typically display a blackbody continuum with a characteristic temperature between $10^{4} - 10^{5}\mathrm{K}$, whilst also showing little temperature evolution post-peak. Additionally, the photospheres of these TDEs appear to decrease in radius over their evolution \citep{van_velzen_seventeen_2021,hammerstein_final_2023}. Post-peak, their luminosities approximately follow a $t^{-5/3}$ power law decay, consistent with the predicted rate of mass fallback from a tidally disrupted star \citep{rees_tidal_1988,phinney_manifestations_1989,gezari_tidal_2021}.

Whilst some of the optically selected TDEs in the samples from \citet{van_velzen_seventeen_2021,hammerstein_final_2023,yao_tidal_2023} are X-ray bright in addition to displaying UV/optical emission, some do not produce detectable X-ray emission and yet are still optically bright. The exact physical source of this UV/optical emission in the absence of detected X-ray emission is a topic of debate, with several hypotheses proposed to explain these observations. These include energy liberated from collisions of in-falling stellar debris rather than accretion \citep{kim_stream-stream_1999}, and reprocessing of X-ray emission from accretion by an extended optically thick outflow \citep{parkinson_optical_2022}.

Sample studies of TDEs have recently begun to reveal sub-populations of TDEs with peculiar characteristics. For example, \citet{hammerstein_final_2023} found 4 events in their sample of 30 which are 2 to 4 times more luminous than any TDEs of other spectral classes in their sample, with $L_{\rm peak} \geq 10^{45}$erg\,s\textsuperscript{-1}. They also appear to display no broad UV/optical emission features, in contrast to most other TDEs previously seen. \citet{hammerstein_final_2023} proposed that they are part of a new distinct rare subclass of extreme `featureless' TDEs. The majority of TDEs appear to show little temperature evolution post-peak \citep{van_velzen_seventeen_2021}. However, a small number of events are characterised by significant post-peak reddening \citep{yao_tidal_2023}, with a notable example being the candidate featureless TDE "Dougie" \citep{vinko_luminous_2014}. In the case of both featureless TDEs and those that display post-peak cooling, the physical mechanism that separates them from `regular' TDEs is currently unknown. Understanding these mechanisms and how these TDEs evolve would be beneficial for sample studies of both SLSNe and TDEs, as some SLSNe can also appear spectroscopically featureless at early times, and all appear to cool post-peak. 

Studies of TDEs at top of the observed luminosity distribution have also begun to find an increasing number of disrupted stars with potentially high masses. In terms of the masses of disrupted stars in TDEs, most events from sample studies such as \citet{hammerstein_final_2023} are estimated to have masses on the order of $\sim1\rm{M_{\odot}}$ or less. As stated by \citet{mockler_weighing_2019}, this is expected due to the stellar Initial Mass Function (IMF) being bottom heavy. However, as the number of TDEs discovered by ongoing sky surveys has grown a small number of events have been discovered that have been hypothesised to be from the disruption of high mass stars ($\gtrsim10\rm{M_{\odot}}$), such as AT2021lwx \citep{subrayan_scary_2023,hinkle_most_2025} and AT2023vto \protect\citep{kumar_at2023vto_2024}. Whilst these types of events are intrinsically rare, their higher luminosity than TDEs of lower mass stars mean that they are observable out to higher redshifts.

In this paper, we present a comprehensive analysis of AT2019cmw, first categorised by \citet{yao_tidal_2023} as a member of the `TDE-featureless' class \citep{hammerstein_final_2023}. It was noted as a prominent outlier in the BTS by \citet{perley_zwicky_2020}. At a redshift of $z = 0.519$, it was the most luminous event seen in the survey by a significant margin. Due to its extreme luminosity and peculiar spectral appearance, it had no analogous events in the BTS, although it is somewhat observationally similar to ASASSN-15lh \citep{dong_asassn-15lh_2016,leloudas_superluminous_2016,mummery_asassn-15lh_2020} and Dougie \citep{vinko_luminous_2014}.

Our extensive follow-up campaign is detailed in Section~\ref{sec:Observations}, with blackbody characteristics derived from multi-wavelength photometry summarised in Section~\ref{sec:Blackbody}. We go on to discuss a variety of possible physical interpretations in Section~\ref{sec:PhysicalOrigin}. In order to constrain other characteristics of the system such as the mass of the disrupted star, we also model the transient using the `cooling envelope' TDE model fit using the Bayesian inference transient modelling software \textsc{Redback}, detailed in Section~\ref{sec:Redback} and Section~\ref{sec:RedbackBolometric}. We also consider an alternate `reprocessing-outflow' model in Section~\ref{sec:Outflow}. Discussion of our findings is detailed in Section~\ref{sec:Discussion}. Our conclusions are summarised in Section~\ref{sec:Conclusions}.

We assume a Flat Lambda-CDM cosmology with ${{\Omega}_m} = 0.3111$, ${{\Omega}_\Lambda} = 0.6889$ and $h = 0.6766$ \citep{aghanim_planck_2020}. Optical and UV magnitudes are in the AB system, and uncertainties are quoted at the $1\sigma$ level unless otherwise specified.

\section{Observations}
\label{sec:Observations}

AT2019cmw was first reported to the Transient Name Server (TNS) on MJD 58567.51 by the ZTF collaboration at an $r$-band magnitude of 19.33 \citep{nordin_ztf_2019}. The transient rose to peak approximately 20 to 30 days after the initial alert from ZTF. Initially flagged as an SLSN candidate due to its blue colour, slow evolution and faint host, a multi-observatory photometric and spectroscopic follow-up campaign was undertaken over the following months. Interest in AT2019cmw was further spurred when spectroscopic observations, as detailed in Section~\ref{sec:Spectroscopy}, revealed it to have a notably high redshift and thus luminosity.

\begin{figure}
\begin{center}
    \includegraphics[width=1\columnwidth]{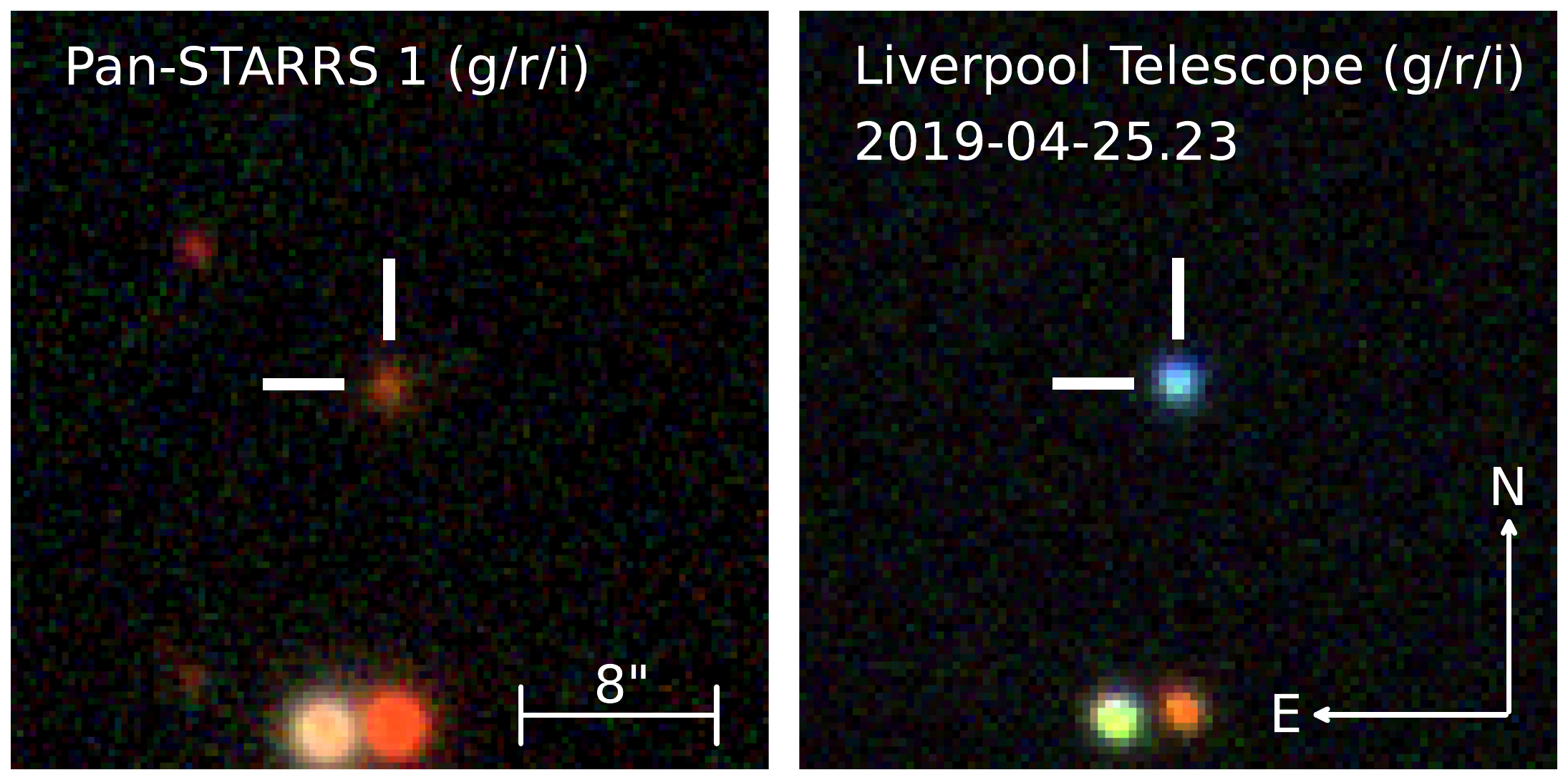}
\end{center}
\caption{\label{fig:HostRGB} \textbf{Left:} PS1 $gri$ composite image of AT2019cmw's host. \textbf{Right:} LT IO:O $gri$ composite image taken $\sim4.8$ days post-peak in AT2019cmw's rest-frame.}
\end{figure}

\begin{figure}
\begin{center}
    \includegraphics[width=1\columnwidth]{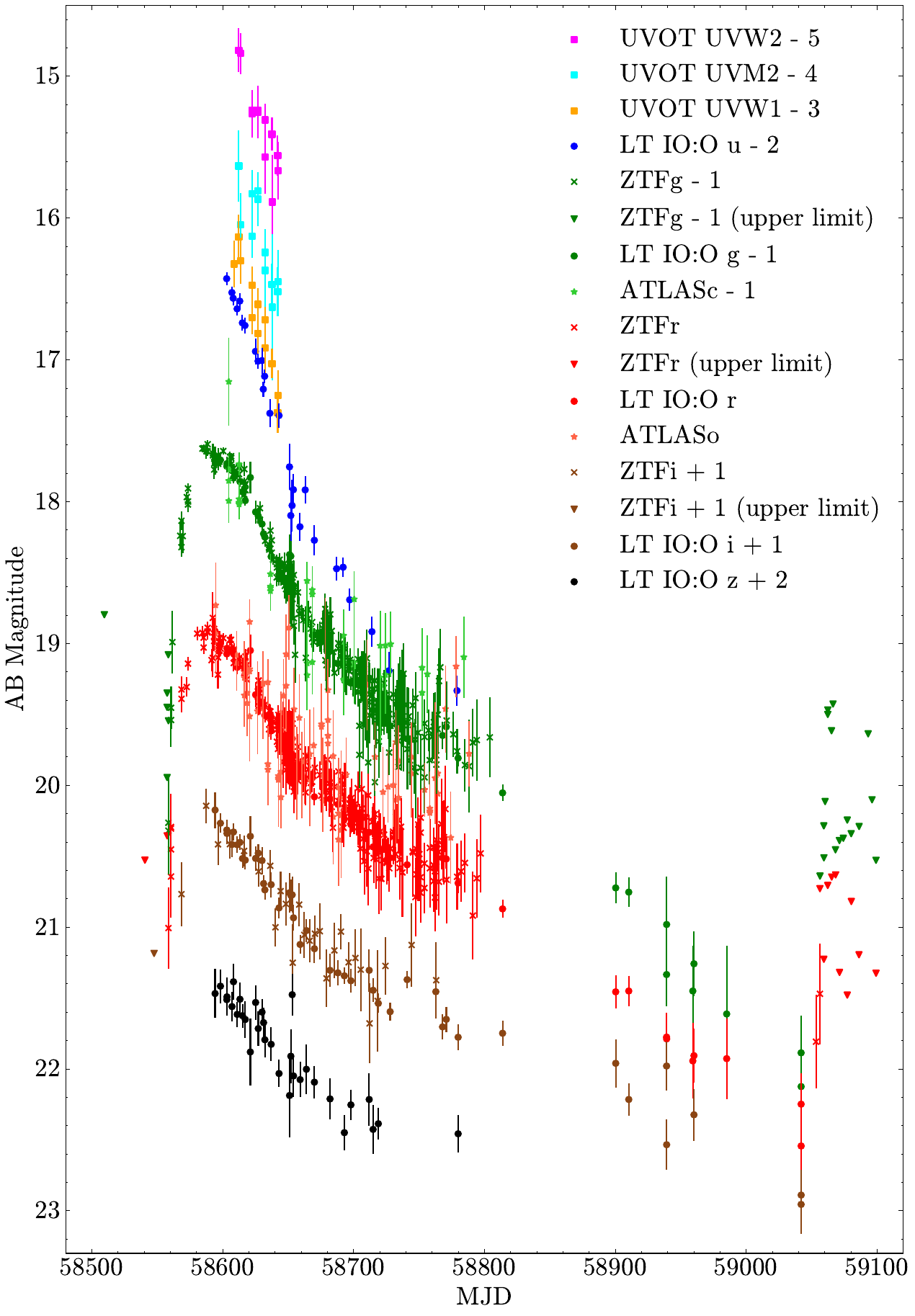}
\end{center}
\caption{\label{fig:Photometry} A combined plot of $ugriz$ photometry from LT IO:O, ZTF $gri$ forced photometry, ATLAS $co$ forced photometry, and $uvw1$, $uvm2$ and $uvw2$ UVOT photometry from \textit{Swift}. An offset has been applied to all photometric bands except r in order to show each band separately. ZTF and LT $gri$ photometry, as well as ATLAS $co$ photometry has been overlaid. ZTF forced photometry $3\sigma$ upper limits have also been plotted up to 50 days before and after the first and last significant forced photometry detections respectively.}
% It seems that you have to add the \protect command when putting citations inside Figure captions
% when using the MNRAS style package
\end{figure}

\subsection{ZTF and ATLAS forced photometry}
\label{sec:ForcedPhotometry}

Forced photometry was carried out using the ZTF forced-photometry service \citep{masci_zwicky_2019} via the ZTF \textsc{Fritz} marshal, an instance of \textsc{SkyPortal} \citep{van_der_walt_skyportal_2019,coughlin_data_2023}, from MJD 58198.42 onwards. Data points with a SNR < 3 were discarded, as well as 41 exposures from field number 1801 due to the reference image for this field being contaminated with flux from the transient. The earliest $3\sigma$ forced photometry detection was at MJD 58558.47 in $g$-band, 9.03 days earlier than the first ZTF alert was reported to TNS, and the last detection was at MJD 59056.34 in $r$-band. Outside of these dates, we find upper limits of $\gtrsim21$ mag throughout.

Forced photometry was also carried out using the ATLAS forced photometry web service \citep{tonry_atlas_2018,smith_design_2020,shingles_release_2021} from MJD 57233.35 onwards. The transient's first $3\sigma$ detection was at MJD 58594.61 and the last detection was at MJD 58788.26 in $o$-band, with upper-limits of $\gtrsim20$ outside of these dates. All datapoints that passed the SNR cut were used, except for one ATLAS $o$-band data point at MJD 58810.19. This detection was $\sim2$ magnitudes brighter than previous ATLAS forced photometry detections, as well as ZTF $r$-band and LT $r$-band detections at similar epochs, and so was considered an outlier and discarded. Reference images used in the image subtraction were updated on MJD 58417 and MJD 58882, before the first significant detection for the former and after the last significant detection for the latter, and so flux contamination from the transient was not an issue.

\subsection{Liverpool Telescope}
\label{sec:LTPhotometry}

AT2019cmw was observed with the Liverpool Telescope (LT) IO:O imager in the Sloan Digital Sky Survey (SDSS) filters $griz$ starting at MJD 58594.14, shortly after the lightcurve's peak in ZTF, with $u$-band observations included from MJD 58603.10. $u$-band observations continued until MJD 58779.86, and $z$-band continued until MJD 58779.87, after which the transient was too faint in these bands. $gri$ observations continued until MJD 59041.99. All images were reduced using the LT automatic data reduction pipeline.

$ugriz$ images taken on the same night were stacked using \textsc{SWarp} \citep{bertin_terapix_2002}. Image subtraction and Point Source Function (PSF) photometry was then performed on $griz$ images using the pipeline developed by Hinds and Taggart et al. (in prep.). Their pipeline uses Point Spread Function Extractor (\textsc{PSFEx}; \citealt{bertin_psfex_2013}) to cross-convolve the PSF from the science image and a PS1 reference image, with the convolved PSF then fit to the science and reference images to measure each respective image. This PSF-fitting method is based on the method from \citet{gal-yam_supernovae_2008} and \citet{fremling_ptf12os_2016}. Image subtraction was then performed using the reference images from the PS1 survey. The magnitudes of comparison stars in the image used to calculate the zeropoint, as well as the magnitude of the transient, were found by measuring fluxes using the extracted PSFs applied to subtracted images.

$u$-band photometry was extracted using relative aperture photometry of the transient and foreground comparison stars in the images, using the python implementation of \textsc{Source Extractor} \citep{bertin_sextractor_1996} \textsc{sep} \citep{barbary_sep_2016}. As the host is red in PS1 with a $g$-band magnitude of $>22$ (see Figure~\ref{fig:HostRGB} and Section~\ref{sec:Host}), we assume that any host contamination within the aperture in $u$-band is negligible compared with the transient flux, and can be ignored. As a result, we do not perform image subtraction for our $u$-band images. The zeropoint of a chosen reference exposure of the transient's field was found using exposures of the LT photometric standard SA114-654 taken on the same photometric night of 2019-09-01. Two observations of the standard at varying airmasses during the night were used to find the $u$-band atmospheric extinction coefficient, and the above-atmosphere filter zeropoint for the night. This zeropoint was used to calculate the above-atmosphere magnitudes of foreground stars in the field, which were then used to find the zeropoints of all epochs. After conducting a median background subtraction, photometry of AT2019cmw was performed using a 1.75" radius aperture.

We also present supplementary LT $griz$ photometry for the featureless TDE AT2018jbv from \citet{hammerstein_final_2023}, which will be discussed in Section~\ref{fig:BBfits} and Section~\ref{sec:TDEcomparison}, taken between MJD 58547.047 and MJD 59024.934. This photometry was reduced using the same method of image subtraction and PSF photometry as detailed above for AT2019cmw's $griz$ photometry. We list LT photometry for AT2018jbv in Table~\ref{tab:AT2018jbv_phot_table}.

\subsection{Swift}
\label{sec:Swift}

Ten epochs of observations were taken using the Neil Gehrels \textit{Swift} Observatory \citep{gehrels_swift_2004} Ultraviolet Optical Telescope (UVOT; \citealt{roming_swift_2005}) and X-Ray Telescope (XRT; \citealt{burrows_swift_2005}) from MJD 58608.48 to MJD 58673.28.

We used the \textsc{uvotsource} package as part of the \textsc{HEASoft} software bundle to reduce UVOT images, which utilises \textsc{ftools} \citep{blackburn_ftools_1999} to extract magnitudes. A 5" radius aperture was chosen to maximise the flux from the transient in order to achieve an optimal SNR. Six circular regions surrounding the target that contained few or no visible sources with radii ranging from 8" to 32" were chosen to compute the background. As was the case for LT IO:O $u$-band photometry as described in Section~\ref{sec:LTPhotometry}, host contamination within the aperture can be considered to be negligible, and so no image subtraction was performed. Photometry from the UVOT $U$, $B$ and $V$ bands was excluded due to poor SNR. Our first and last significant detections were at MJD 58608.54 and MJD 58642.20 respectively, both in $UVW1$.

XRT observations were taken simultaneously with UVOT observations for a total of $\mathrm{14.94\,ks}$. A stack of all 10 epochs yielded a non-detection with a $3\sigma$ count-rate upper limit of $<5.355\times10^{-4}$ ct s\textsuperscript{-1} at the transient's position. Assuming a spectral shape of a power law with a photon index of 2, and with a Galactic neutral hydrogen column density of $4.33\times10^{20}$cm\textsuperscript{-2} \citep{hi4pi_collaboration_hi4pi_2016} whilst ignoring any absorption, this corresponds to an absorbed flux upper limit of $<1.883\times10^{-14}$ erg\,s\textsuperscript{-1}cm\textsuperscript{-2} between 0.3-10 keV.

\subsection{NEOWISE}
\label{sec:NEOWISEPhotometry}

Photometry was obtained using the NASA IRSA online catalogue from the Near-Earth Object Wide-field Infrared Survey Explorer (NEOWISE) Reactivation Mission \citep{mainzer_initial_2014}, an infrared sky survey mission using the $3.4{\mu}m$ (W1) and $4.6{\mu}m$ (W2) filters from the Wide-Field Infrared Explorer (WISE; \citealt{wright_wide-field_2010}). Using IRSA, a search was performed at the target's location within a circular region with 2.5" radius, with the resulting photometry then binned over intervals of 100 days. For the filter W1 only detections with a SNR > 2 were used, and for W2 all detections were used as detections in this filter were more marginal. An error weighted average of these fluxes was then used to calculate the magnitudes for each 100-day bin.

\begin{figure}
\begin{center}
    \includegraphics[width=1\columnwidth]{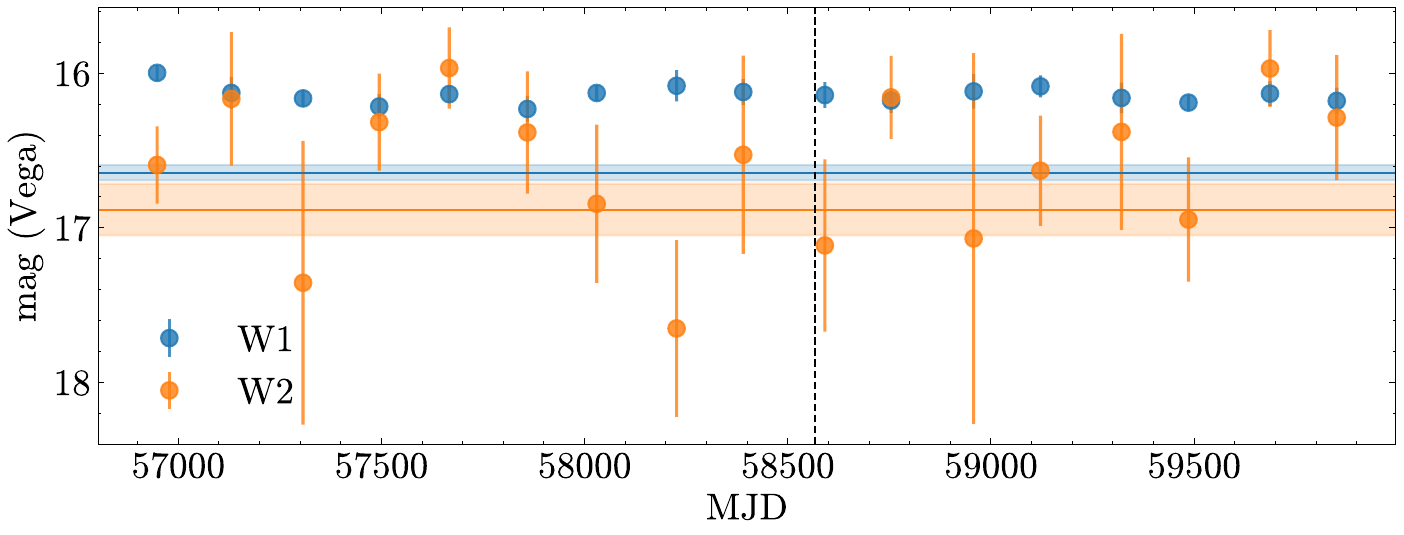}
\end{center}
\caption{\label{fig:NEOWISE} Vega-magnitude plot of NEOWISE W1 (blue) and W2 (orange) photometry at the location of AT2019cmw from MJD 56948.07 to MJD 59850.88, with AT2019cmw's earliest ZTF forced photometry detection at MJD 58558.47 marked by the black vertical dashed line. Datapoints show NEOWISE photometry values found using the method outlined in Section~\ref{sec:NEOWISEPhotometry}, with error bars corresponding to $1\sigma$ errors. Solid horizontal lines and shaded regions show ALLWISE W1 and W2 photometry values in the same colours.}
\end{figure}

Figure~\ref{fig:NEOWISE} shows the NEOWISE photometry found using this method from MJD 56948.07 to MJD 59850.88 for AT2019cmw's host. To search for any IR variability from AT2019cmw's host during this time period that may be caused by the transient or a pre-existing AGN, we compare flux from individual epochs of NEOWISE photometry to the weighted averages of all epochs for both filters. Most epochs show no significant deviation from the weighted mean in each filter. The most significant deviation that we see is at MJD 56948.03, it is brighter than the weighted mean in W1 with a deviation of $\sim2.6\sigma$, whereas in W2 it is fainter with a deviation of $\sim0.8\sigma$. As a $\sim2.6\sigma$ is expected $\sim1\%$ of the time for any particular measurement, there is a $\sim34\%$ chance of this occurring for a photometric data point over our observations. Also, this deviation did not occur in both filters simultaneously. We therefore do not consider the location of AT2019cmw to to have significantly variable IR flux during the time period in which it was observed.

By subtracting the average baseline flux pre-transient from the brightest detections at the location of AT2019cmw post-transient onset, at MJD 59121.73 for W1 and MJD 59686.59 for W2, we find upper-limits of W1 > 16.74 and W2 > 16.27.

We also plot ALLWISE photometry from MJD 55302.78 to MJD 55487.19 for comparison, which has magnitudes of $W1 = 16.64\pm0.05$ and $W2 = 16.88\pm0.17$ (Vega). The host was not significantly detected in $W3$ or $W4$. As can also be seen in Figure~\ref{fig:NEOWISE}, AT2019cmw's NEOWISE photometry in W1 is consistently significantly brighter than in ALLWISE at all observed epochs. However, we do not consider this discrepancy to be physical, as for faint objects such as AT2019cmw's host NEOWISE is known to overestimate flux in the W1 band compared to ALLWISE by up to half a magnitude\footnote{\url{https://wise2.ipac.caltech.edu/docs/release/neowise/expsup/sec2_1c.html}}.

\subsection{Spectroscopy}
\label{sec:Spectroscopy}

\begin{figure*}
\begin{center}
    \includegraphics[width=2\columnwidth]{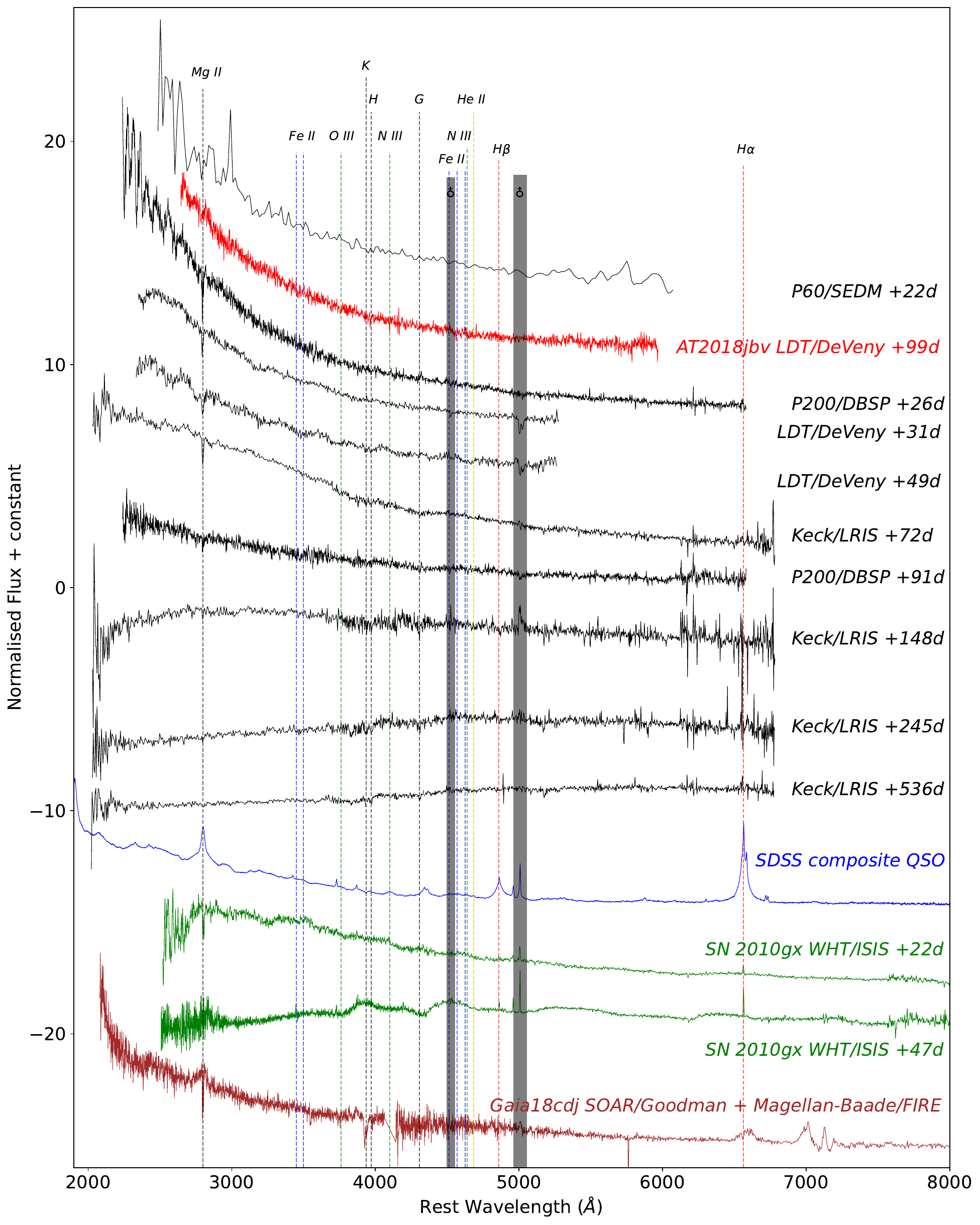}
\end{center}
\caption{\label{fig:Spectra_cmw} Spectra of AT2019cmw from Palomar P60, Palomar P200, LDT and Keck LRIS in descending order of their observation, with time in days since first ZTF forced photometry detection in the AT2019cmw's rest frame. The positions of key spectral lines from the classification scheme outlined by \protect\citet{van_velzen_seventeen_2021} have been labelled. $\mathrm{Mg II}$ absorption and the K, H and G Fraunhofer lines of calcium have also been labelled. Late-time LRIS spectra are host galaxy dominated. Also plotted for comparison are an LDT spectrum of the featureless TDE AT2018jbv \protect\citep{hammerstein_final_2023}, an SDSS QSO composite spectrum \protect\citep{vanden_berk_composite_2001} and two WHT ISIS spectra of the SLSN-I SN\,2010gx \protect\citep{pastorello_ultra-bright_2010}. We also show a combined SOAR Goodman and Magellan-Baade FIRE spectrum of the ENT Gaia18cdj \protect\citep{hinkle_most_2025}, taken $\sim2$ years and $\sim3.5$ years post-first detection in the transient's rest-frame respectively. Spectra have been normalised to their respective mean rest-frame flux between $\mathrm{4000\,{\AA}}$ and $\mathrm{5000\,{\AA}}$.}
% It seems that you have to add the \protect command when putting citations inside Figure captions
% when using the MNRAS style package
\end{figure*}

% Please add the following required packages to your document preamble:
% \usepackage{graphicx}
\begin{table*}
\centering
\caption{Spectroscopic setups for observations shown in Figure~\ref{fig:Spectra_cmw}.}
\label{tab:SpecTable}
\begin{tabular}{lccccc}
 MJD      &  Phase            &  Telescope/Instrument &  Grating           &  Slit width &  Exposure  \\ \hline
         &  d, rest &                      &                   &  "          &  s         \\ \hline
 58591.38 &  +22              &  P60/SEDM             &  IFU               &  N/A        &  2250      \\
 58597.49 &  +26              &  P200/DBSP            &  600/4000          &  1.5        &  600       \\
 58605.45 &  +31              &  LDT/DeVeny           &  300/4000          &  1.5        &  1200      \\
 58632.44 &  +49              &  LDT/DeVeny           &  300/4000          &  1.5        &  1600      \\
 58668.54 &  +72              &  Keck/LRIS            &  400/3400,400/8500 &  1          &  900,800   \\
 58696.28 &  +91              &  P200/DBSP            &  600/4000          &  1.5        &  900       \\
 58783.31 &  +148             &  Keck/LRIS            &  400/3400,400/8500 &  1          &  600,540   \\
 58930.61 &  +245             &  Keck/LRIS            &  400/3400,400/8500 &  1          &  900,900   \\
 59372.59 &  +536             &  Keck/LRIS            &  400/3400,400/8500 &  1          &  1200,1130
\end{tabular}
\end{table*}

Spectra were taken using the SED Machine (SEDM; \citealt{blagorodnova_sed_2018,rigault_fully_2019}) on the Palomar 60-inch telescope, the Double Spectrograph (DBSP; \citealt{oke_efficient_1982}) on the Palomar 200-inch telescope, the DeVeny Optical Spectrograph \citep{bida_first-generation_2014} on the 4.3 meter Lowell Discovery Telescope (LDT), and the Low Resolution Imaging Spectrometer (LRIS; \citealt{oke_keck_1995}) on the 10 meter Keck I telescope. These spectra were reduced using the same methods detailed in Appendix B of \citet{yao_tidal_2022}. Instrumental setups for each observation are detailed in Table~\ref{tab:SpecTable}.

As seen in Figure~\ref{fig:Spectra_cmw} AT2019cmw remains spectroscopically featureless for its entire observed evolution in the optical, and displays none of the prominent broad emission features seen in most TDEs. For instance, the hydrogen, helium, oxygen and nitrogen emission features used to define the classification scheme from \citet{van_velzen_seventeen_2021} are absent. Less common features such as Fe II, as seen in the case of AT2018fyk \citep{wevers_evidence_2019}, are also not present. At late times, spectra appear host-dominated.

As is also shown in Figure~\ref{fig:Spectra_cmw}, $\mathrm{Mg II}$ absorption from the intervening interstellar/intergalactic medium can be seen in the majority of our spectra. We also see the $\mathrm{K}$, $\mathrm{H}$ and $\mathrm{G}$ Fraunhofer lines of calcium in many of our spectra at the same redshift. Using these lines, we measure a redshift for AT2019cmw of $z = 0.519$. After applying a $2.5{\rm{log}_{10}}(1+z)$ $K$-correction, this implies a peak rest-frame $u$-band magnitude of M = -23.6, as was also found by \citet{perley_zwicky_2020}.

\subsection{VLA}
\label{sec:VLA}

The transient's location was observed with the Karl G. Jansky Very Large Array (VLA; \citealt{thompson_very_1980,perley_expanded_2011}) radio telescope as part of a late-time TDE follow-up program (ID: 23A-280, \citealt{yao_optically_2025}). The observation of AT2019cmw started at MJD 60033.34, 1474.86 days after the first ZTF forced photometry detection. The total on-source integration time was $\mathrm{5.026\,hrs}$ in the VLA C-band (4-8 GHz), with array configuration B used. AT2019cmw was undetected down to a $3\sigma$ upper limit of \(4.5 {\mu}\rm{Jy}\). The data was analysed following the standard radio continuum image analysis procedures in the Common Astronomy Software Applications (\textsc{CASA}; \citealt{team_casa_2022}).

\section{Characteristics}
\label{sec:Characteristics}

In this section, we use the data described in Section~\ref{sec:Observations} to infer physical properties of AT2019cmw.

\subsection{Rise time to peak}
\label{sec:RiseTime}

In order to constrain the date of peak brightness, we fit curves to the first 100 days of $g$-band forced photometry using bootstrapped Locally Weighted Scatterplot Smoothing (LOWESS; \citealt{cleveland_robust_1979,cleveland_lowess_1981}), incorporating the \textsc{interp1d} function from \textsc{SciPy} \citep{virtanen_scipy_2020}. The local regression was repeated using a range of fractions of the data from 0.07 to 0.17 in increments of 0.01, with each fit repeated 1000 times. From this we estimate that the transient peak occurred on MJD $58590.9^{\textsuperscript{$+$2.3}}_{\textsubscript{$-$2.7}}$ in observer-frame $g$-band, which we use in the analysis hereon. AT2019cmw thus has a rise time from first significant detection to peak brightness of $21.35^{\textsuperscript{+1.51}}_{\textsubscript{-1.78}}$ days in its rest frame.

\subsection{Pseudo-blackbody fits}
\label{sec:Blackbody}

\begin{figure}
\begin{center}
    \includegraphics[width=1\columnwidth]{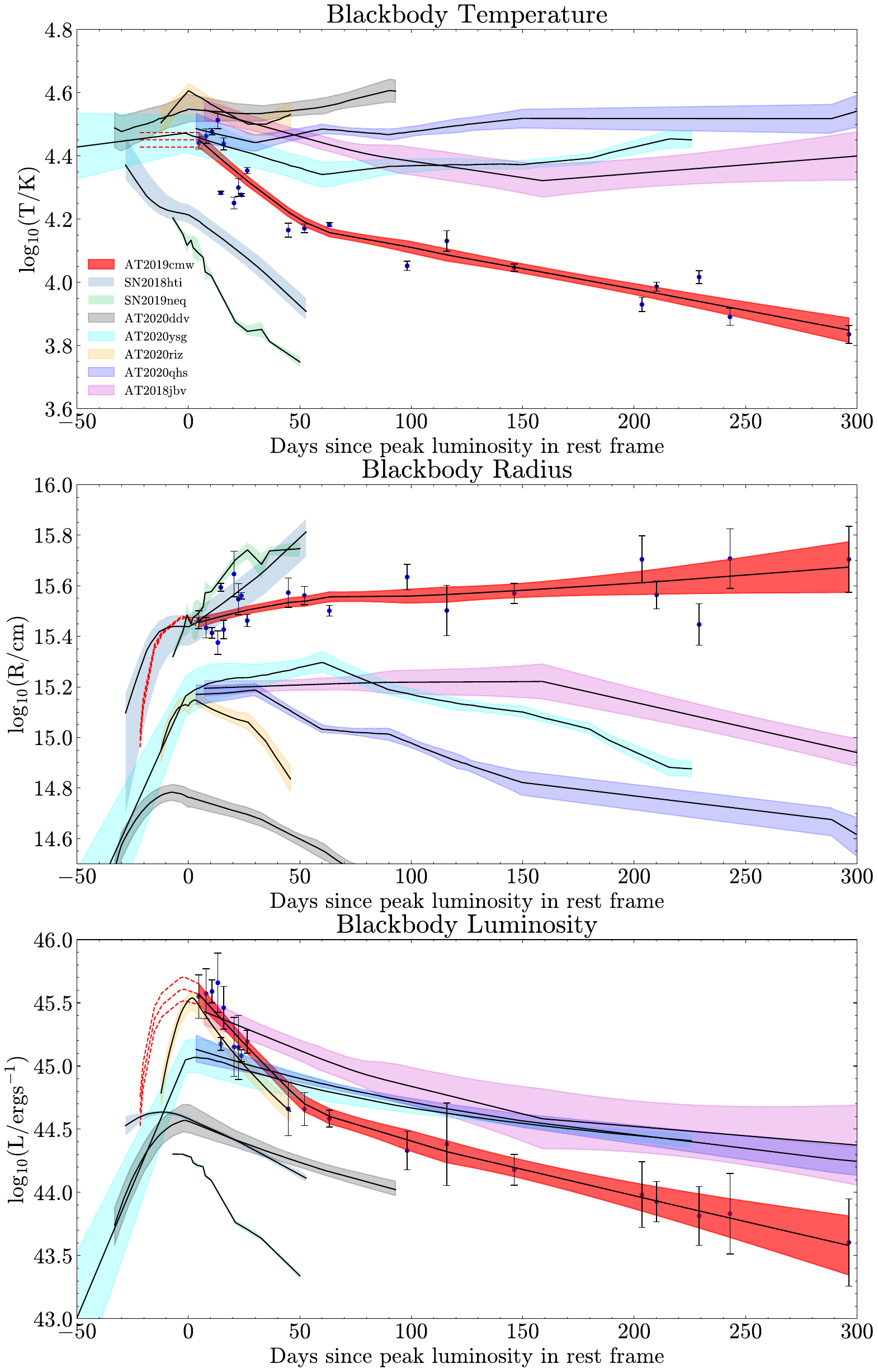}
\end{center}
\caption{\label{fig:BBfits} Pseudo-blackbody temperatures (\textbf{top}), radii (\textbf{middle}) and luminosities (\textbf{bottom}) of AT2019cmw and four featureless TDEs (AT2018jbv, AT2020riz, AT2020qhs and AT2020ysg) from \protect\citet{hammerstein_final_2023}, with the most luminous TDE with spectral features from their sample (AT2020ddv, classified as TDE-He) also included for comparison). Values for AT2018jbv were derived using the method described in Section~\ref{sec:Blackbody}, with the closest to peak modelled epoch ($\sim7$ days post-peak) epoch used an an initial value. The SLSNe-I events SN2018hti and SN2019neq from \protect\citet{chen_hydrogen-poor_2023} are also displayed. SN2018hti's data has been smoothed for better visual clarity. One data point for AT2019cmw fit using LT $griz$ and \textit{Swift} UV data $\sim14.5$ days post-peak is significantly deviating from the general trend in both temperature and radius.}
\end{figure}

We fit blackbody SEDs to AT2019cmw's ZTF, ATLAS, LT and \textit{Swift} photometry. Optical photometry of AT2019cmw was corrected for Galactic foreground extinction using $A_{\lambda}/E(B-V)$ values derived by \citet{schlafly_measuring_2011}, and \textit{Swift} UVOT UV photometry was corrected using $A_{\lambda}/E(B-V)$ values from Table 2 of \citet{code_empirical_1976}. All photometry was corrected using $E(B-V) = 0.0438$ mag as measured by \citet{schlafly_measuring_2011} at the location of the transient. Effective wavelengths for each observed filter were found using SVO filter profile service \citep{rodrigo_svo_2012, rodrigo_svo_2020}. Photometry was then corrected for time dilation and redshift effects.

In order to determine the radius and temperature evolution of AT2019cmw, we then fit Planck function SEDs using the \textsc{SciPy} \textsc{curve\_fit} function to our photometry in 1 day intervals. We only consider days with observations in more than 3 photometric bands, as we are simultaneously fitting two free parameters. The luminosity at each epoch was then found using \(L = 4{\pi}R^{2}{\sigma}T^{4}\).

$1\sigma$ confidence intervals for AT2019cmw's blackbody characteristics were deduced using the bootstrapping method. Using the \textsc{resample} function from \textsc{Scikit-learn} \citep{buitinck_api_2013}, the photometry and the uncertainties on each measurement were randomly resampled a large number of times (>1000) in order to produce a distribution of derived values of $L$, $T$ and $R$ for each fitted epoch. The overall trends of its blackbody evolution were then found by interpolating its derived blackbody characteristics using bootstrapped LOWESS.

Due to poor photometric coverage, we were unable to constrain its pre-peak blackbody characteristics using the method outlined above. To obtain estimates of its blackbody evolution pre-peak, we make the assumption that AT2019cmw's cooling began at peak and fix its temperature to the peak value found by LOWESS interpolation of post-peak epochs as detailed above. An approximately steady blackbody temperature pre-peak has been previously observed in a variety of TDEs. For example, Figure 10 from \citet{holoien_discovery_2019} shows this behaviour for the TDEs ASASSN-19bt and ASASSN-18pg \citep{holoien_rise_2020,leloudas_spectral_2019}. The featureless TDEs from \citet{hammerstein_final_2023}, as can be seen in Figure~\ref{fig:BBfits}, also showed little temperature evolution pre-peak. Additionally, this has been shown to be the case for TDEs that cool post-peak. As can be seen in Supplementary Figure 2 from \citet{angus_fast-rising_2022}, the cooling TDE AT2020neh did not appear to show significant temperature evolution until after peak luminosity, although its coverage is poor. We use this assumption, as AT2019cmw has previously been classified and analysed as a TDE by \citet{yao_tidal_2023} and \citet{mummery_fundamental_2024}, with our own arguments for this interpretation presented later in Section~\ref{sec:PhysicalOrigin}. We note that although we use these pre-peak data points for our models in Section~\ref{sec:RedbackBolometric} and Section~\ref{sec:Outflow}, if this assumption is not physically accurate for AT2019cmw it does not materially affect our overall conclusions.

Figure~\ref{fig:BBfits} shows the evolution of AT2019cmw's temperature, radius and luminosity over time obtained using the above methods. As can be seen in Figure~\ref{fig:BBfits}, AT2019cmw has a temperature of $\mathrm{\sim10^{4.45}\,K}$ close to peak, and then steadily cools over the next $\sim50$ days. It then continues cooling at a slower rate to $\mathrm{\sim10^{3.85}\,K}$ at $\sim300$ days post-peak. Its radius appears to gradually increase/plateau post-peak, increasing from $\mathrm{\sim10^{15.45}\,cm}$ to $\mathrm{\sim10^{15.65}\,cm}$. Its luminosity peaks at $\mathrm{\sim10^{45.6}\,erg\,s^{-1}}$. The rate of luminosity decline broadly follows its temperature evolution, steadily declining for the first $\sim50$ days before declining at a slower rate for the following $\sim250$ days to $\mathrm{\sim10^{43.6}\,erg\,s^{-1}}$. By integrating our derived luminosities over time, we find that AT2019cmw emitted $\sim1.7\times10^{52}\,\mathrm{erg}$ between its first detection and our last modelled epoch at MJD 59041. This value was calculated assuming no host extinction and excludes emission outside of the period during which it was significantly detected, and so is a lower limit on the true total radiated energy. The derived temperatures, radii and luminosities for each fitted epoch of AT2019cmw's photometry are listed in Table~\ref{tab:AT2019cmw_blackbody_table}.

In Figure~\ref{fig:BBfits} we compare physical parameters derived from pseudo-blackbody SED fits for AT2019cmw to the blackbody evolution of 4 spectroscopically similar featureless TDEs, as well as the TDE-He event AT2020ddv, from the sample of \citet{hammerstein_final_2023}. Temperatures, radii and luminosities for AT2020riz, AT2020qhs, AT2020ysg and AT2020ddv were taken from \citet{hammerstein_final_2023}. For AT2018jbv, we apply the same method we used for AT2019cmw to its supplementary LT photometry detailed in Section~\ref{sec:LTPhotometry}, alongside ZTF, ATLAS and \textit{Swift} photometry from \citet{hammerstein_final_2023}. We list AT2018jbv's blackbody parameters derived using this method in Table~\ref{tab:AT2018jbv_blackbody_table}. We only consider these 4 featureless TDEs, as although other featureless TDEs have been catalogued by sample studies such as \citet{yao_tidal_2023}, they lack the comprehensive post-peak blackbody modelling needed to make the same comparison.

\subsubsection{Possible deviation from blackbody}
\label{sec:UVSuppression}

\begin{figure}
\begin{center}
    \includegraphics[width=0.843\columnwidth]{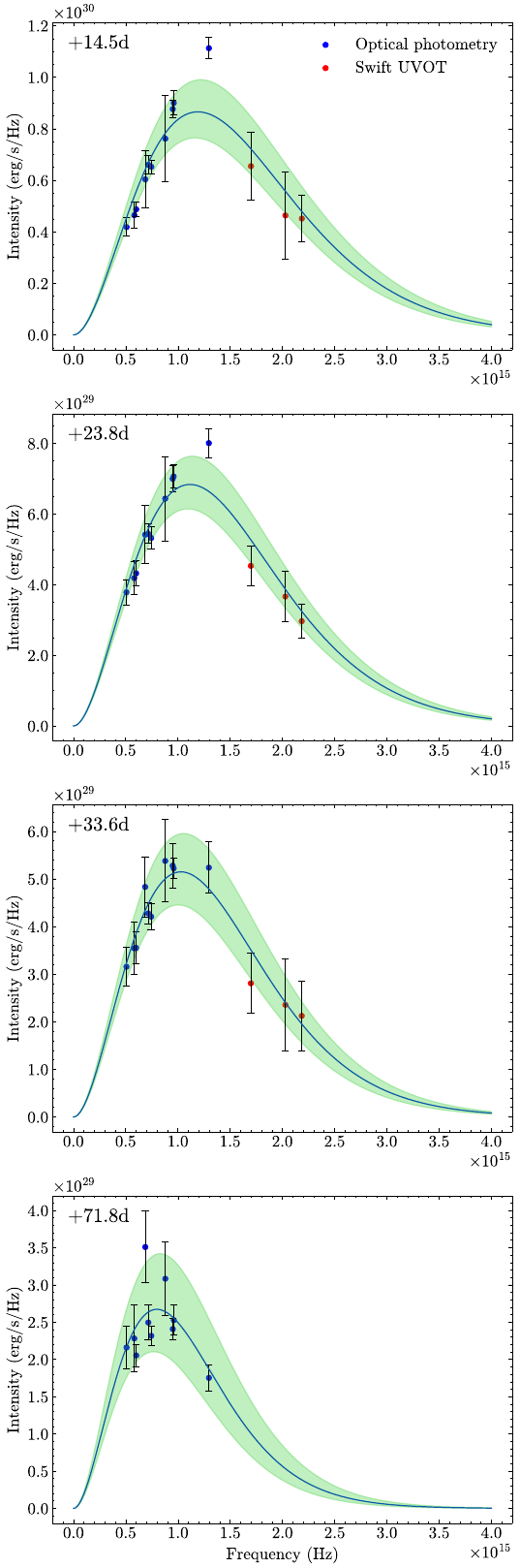}
\end{center}
\caption{\label{fig:SEDs} Blackbody SED fits to interpolated rest-frame multi-wavelength photometry for 14.5 days, 23.8, 33.6 days and 71.8 days post-peak luminosity in rest-frame. Fits are in chronological order from top to bottom. Optical and \textit{Swift} UVOT photometry are indicated by blue and red points respectively. The shaded region above and below the curve represents the $1\sigma$ confidence interval of our fit.}
\end{figure}

As can be seen in Figure~\ref{fig:BBfits}, we find a significantly lower blackbody temperature and higher blackbody radius for an epoch $\sim14.5$ days post-peak compared with optical-only fits at a similar time. In Figure~\ref{fig:SEDs} we show blackbody SEDs fit to snapshots of interpolated photometry at the beginning, middle, and end of the period during which \textit{Swift} UV data was gathered between 14.5 and 33.6 days post-peak in order to probe this further. We also show an SED-fit at 71.8 days post-peak to highlight the continuing evolution of AT2019cmw's SED. At 14.5 days post-peak luminosity in rest-frame, the \textit{Swift} UV bands are underluminous compared with other bands at optical wavelengths assuming a blackbody SED. At 33.6 days post-peak, all bands appear consistent with a blackbody SED. 

In case this behaviour was due to the presence of multiple emitting regions with differing temperatures and radii shortly post-peak, we attempted a multi-component blackbody fit for the epochs detailed above. However, this did not improve the quality of our fits, suggesting that a single component is dominant in the optical/UV.

We discuss this potential UV underluminosity further in Section~\ref{sec:UVBehaviour}.

\subsection{Post-peak colours}
\label{sec:Colours}

\begin{figure}
\begin{center}
    \includegraphics[width=1\columnwidth]{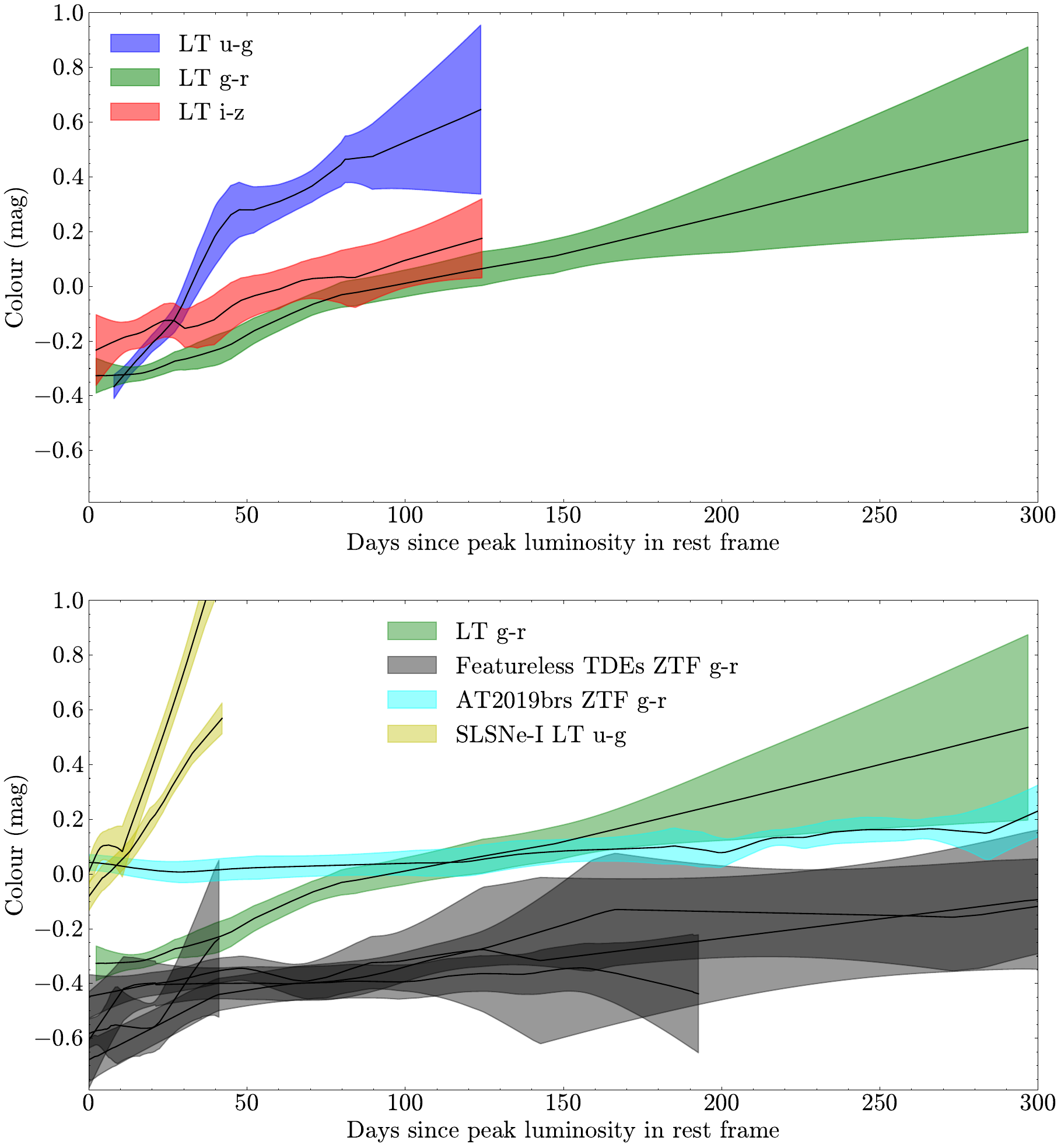}
\end{center}
\caption{\label{fig:Colours} \textbf{Top:} Observed LT $u-g$, LT $g-r$ and LT $i-z$ colours for AT2019cmw in the first 300 days since peak luminosity in its rest-frame. \textbf{Bottom:} LT $g-r$ colour of AT2019cmw plotted against ZTF $g-r$ colours of the 4 featureless TDEs presented by \protect\citet{hammerstein_final_2023} and the luminous ANT AT2019brs from \protect\citet{wiseman_systematically_2025} since peak luminosity in each transient's respective rest-frame. For comparison we also show LT $u-g$ colours of the SLSNe-I events SN\,2018hti and SN\,2019neq from \protect\citet{chen_hydrogen-poor_2023}. As these events were at redshifts of $z = 0.063$ and $z = 0.1075$ respectively, their $u-g$ colours provide a more accurate rest-frame comparison than their $g-r$ colours to the plotted higher redshift TDEs and ANTs at $z\sim0.3-0.5$. Shaded regions show errors propagated from Lowess interpolated photometry.}
\end{figure}

As seen in Figure~\ref{fig:Colours}, AT2019cmw has a blue colour of $<0$ mag close to peak across broadband filter pairs between u-band and z-band. It then begins to redden with time, with its $u-g$ colour increasing significantly faster than colours between other observed optical bands in the first $\sim50$ days post-peak, before slowing to a rate more comparable with these other bands. As they appear spectroscopically similar and have comparable peak luminosities, we compare AT2019cmw's $g-r$ colour with the 4 featureless TDEs from \citet{hammerstein_final_2023} in the first 300 days post-peak in each transient's respective rest frame. Whilst they arguably show some colour evolution, none of these other events show the same significant change that AT2019cmw does, and maintain a colour of $\leq0$ throughout their evolution. AT2019cmw by comparison reddens in its LT photometry from a $g-r$ colour of $\sim-0.35$ mag near-peak to $\sim0.50$ mag at late times. Compared with the luminous ANT AT2019brs \citep{wiseman_systematically_2025} AT2019cmw appears $\sim0.3$ mag bluer at peak, whereas at late times it is comparable in colour. AT2019cmw is also $\sim0.3$ mag bluer than the SLSNe-I SN\,2018hti and SN\,2019neq \citep{chen_hydrogen-poor_2023} at peak, and reddens significantly slower than both post-peak. See Section~\ref{sec:TDEcomparison} for a more detailed comparison of the evolution of inferred blackbody parameters between AT2019cmw and the featureless TDEs from \citet{hammerstein_final_2023}.

\subsection{Host Galaxy}
\label{sec:Host}

$rizy$ PS1 photometry of AT2019cmw's host is listed in \citet{yao_tidal_2023}. The host galaxy is only marginally detected in the Pan-STARRS1 $g$ band image. To recover the magnitude in this band, forced aperture photometry at the host location was performed using a 2" radius, with a small aperture correction (measured using the r-band image) applied. From this, we find a $g$-band magnitude of $g = 22.57\pm0.29$ for AT2019cmw's host. Combined with the photometry from \citet{yao_tidal_2023}, this shows it to be a red galaxy. We show a PS1 $gri$ composite image of AT2019cmw's host in Figure~\ref{fig:HostRGB}.

AT2019cmw's host SED was fit in the sample study from \citet{yao_tidal_2023}, in which they find it to be consistent with being a normal non-active galaxy.

The lack of any detected variability in NEOWISE photometry at the source location from 1081.5 days prior to and 829.5 days post peak luminosity in rest-frame means that any IR-reflecting material that may exist out to $2\times{10^{18}}\,\mathrm{cm}$ from the event, such as a dusty AGN torus, was not significantly detected (using $d=ct$). AT2019cmw may begin to display rising IR emission in the future, as radiation from the event reaches more distant IR-reflecting material. This is not particularly constraining. \citet{jiang_infrared_2021}, for example, detected IR variability in 23 optically selected TDEs, however they were all at significantly lower redshift than AT2019cmw.

We fit a gaussian line profile to the host spectrum at the location of $\mathrm{H\alpha}$ in order to attempt to measure the galaxy's star formation rate (SFR). The FWHM of this gaussian, $6.9\,{\AA}$, corresponds to the FWHM resolution for the LRIS 400/8500 grating using a 1" slit\footnote{\url{https://www2.keck.hawaii.edu/inst/lris/dispersive_elements.html}}. This yielded a $3\sigma$ flux upper limit of $\lesssim3.5\times10^{33}\mathrm{\,erg\,s^{-1}}$. This corresponds to an approximate $3\sigma$ upper limit on the star formation rate of $\lesssim0.3\,{\rm{M_{\odot}yr^{-1}}}$ using the relation from \citet{kennicutt_past_1994}, and a specific star formation rate of $\lesssim3.7\times10^{-12}\,\rm{yr^{-1}}$ for a stellar mass of $10^{10.88}\rm{M_{\odot}}$ \citep{yao_tidal_2023}.

\subsubsection{Host-Transient Offset}
\label{sec:offset}

In order to quantify the offset between the transient and its host galaxy, we compare the measured position of the transient in a host-subtracted LT image with the position of the host in PS1 reference imaging. To do this, we chose a high quality LT $i$-band image with good seeing and low moon contamination from 2019-06-08, and a stacked $i$-band PS1 cutout of the host galaxy. $i$-band images were chosen as the host galaxy has the highest S/N in this band.

To perform the subtraction, the LT image was first aligned with the PS1 reference image using \textsc{SWarp}. PSFs for both images were measured using \textsc{PSFEx} \citep{bertin_psfex_2013}, and each image was convolved with the PSF of the other using \textsc{Source Extractor}. The alignment of the convolved images is refined using \textsc{SciPy}'s \textsc{ndimage.shift} function. We then normalise the images for subtraction, using a chosen subset of foreground reference stars that pass a number of quality cuts.

After subtraction, we measure the difference in the position of the transient and the host in subtracted and reference images using \textsc{Source Extractor}. A final correction to this difference was then applied using the mean of the offset of the foreground reference stars between both images. The error of the position of both the transient in host-subtracted LT images and the host in PS1 was found using the scatter in the position offset of foreground reference stars between both images.

Using this method, an offset of 0.195 PS1 pixels was found between AT2019cmw and its host, and a scatter of 0.449 pixels using foreground stars. This corresponds to an offset of 0.049" and a scatter of 0.112", and thus $0.57\,\mathrm{kpc}$ with a scatter of $1.31\,\mathrm{kpc}$ at a redshift of $z = 0.519$. We do not see any significant offset between the transient and its host-galaxy, and thus consider the transient as nuclear.

\subsection{X-ray and Radio non-detections}
\label{sec:XrayResults}

Our X-ray flux upper limit, as detailed in Section~\ref{sec:Swift}, translates to an X-ray luminosity upper limit of $L_{X}\lessapprox2\times10^{43}$\,erg\,s\textsuperscript{-1} between 0.3-15.2 keV, between $\sim12$ and $\sim54$ days post-peak, in AT2019cmw's rest-frame. This is not particularly constraining on the presence of X-ray emission. For example, in the study by \citet{hammerstein_final_2023} for TDEs with a redshift of $z\leq0.075$ they find 6 TDEs with $L_{X}\leq10^{42}$erg\,s\textsuperscript{-1} and 7 TDEs with $L_{X}\geq10^{42}$erg\,s\textsuperscript{-1}. However, some of the TDEs in their sample considered `X-ray bright' would be detected at AT2019cmw's redshift at the depth of our observations. AT2020ddv, for instance, has a \textit{Swift} XRT X-ray luminosity of $\gtrsim3\times10^{43}$\,erg\,s\textsuperscript{-1} at a similar time to AT2019cmw post-peak. In addition to this, AT2020ddv is almost an order of magnitude less luminous at optical/UV wavelengths than AT2019cmw. At the time of our observations, the ratio between AT2019cmw's derived blackbody luminosity and its X-ray luminosity upper limit was $\sim50-100$. As can be seen in Figure 8 from \citet{hammerstein_final_2023}, this would place it near the upper end of their $L_{BB}/L_X$ distribution for X-ray detected TDEs. In their sample, at a similar time post-peak to our \textit{Swift} XRT observations 6 TDEs had $L_{BB}/L_X$ between $\sim1$ and $\sim50$, and 2 showing $L_{BB}/L_X$ between $\sim100$ and $\sim300$. This suggests that AT2019cmw was not particularly X-ray luminous at the time of our observations.

Assuming a spectral shape of $F_{\nu}\sim{{\nu}^{-1}}$ for our K-correction, our VLA C-band flux upper-limit of $<4.5\,{\mu}$Jy corresponds to a luminosity upper limit of ${\nu}{L}_{\nu} < 2.8\times10^{38}\mathrm{erg\,s^{-1}}$ between 4-8 GHz 1474.86 days after first detection (970.94 days in rest frame). As seen in the sample paper from \citet{cendes_ubiquitous_2024} the majority of optically-selected TDEs are not particularly radio-loud, with many events being detected at luminosities below our detection limit at a similar time post first-detection. However, they show that a subset of TDEs are extremely radio-luminous due to on-axis relativistic jetted emission. Swift J164449.3+573451 \citep{cendes_radio_2021} for instance was detected at a luminosity over two orders of magnitude above our detection limit for AT2019cmw's redshift. This places tight constraints on the presence of any on-axis relativistic jetted emission in AT2019cmw. We discuss this further in Section~\ref{sec:Radiocomparison}.

\section{Physical origin scenarios for AT2019cmw}
\label{sec:PhysicalOrigin}

In this section, we consider several different physical interpretations in order to explain AT2019cmw's observed characteristics.

AT2019cmw has been previously classified and analysed as a featureless TDE in the sample studies from \citet{mummery_fundamental_2024} and \citet{yao_tidal_2023}, and appears spectrally similar to the featureless TDEs from \citet{hammerstein_final_2023}, as we show in Figure~\ref{fig:Spectra_cmw} where AT2018jbv is plotted for comparison. However, it lacks a concrete `smoking gun' that definitively pins down its physical origin, such as detected X-ray or radio emission. As a result, we briefly consider here some potential alternative physical interpretations for AT2019cmw's observed characteristics.

\subsection{Extreme Supernova}
\label{sec:Supernova}

Despite its host's low rate of star formation, as detailed in Section~\ref{sec:Host}, its significant temperature evolution post-peak motivates us to consider an extreme SLSN scenario for AT2019cmw.

The most extreme magnetar spin-down powered models of SLSN-I from \citet{sukhbold_most_2016} generate a peak luminosity of $L = 2\times10^{46}$erg\,s\textsuperscript{-1}, which could possibly explain our observations. However, as seen in Figure~\ref{fig:BBfits}, at peak AT2019cmw is significantly hotter than typical SLSNe. Additionally, SN2020hti was highlighted by \citet{chen_hydrogen-poor_2023} to be a high-temperature outlier in their sample of SLSN-I events, AT2019cmw displays significantly higher temperatures at all comparative epochs. Although its radius evolves slower than the SLSN-I events seen in Figure~\ref{fig:BBfits}, AT2019cmw also displays a plateau in its blackbody radius for at least 300 days in rest frame, which could be explained by the photosphere receding into the expanding ejecta of a supernova as it becomes optically thin. 

However, the strongest line of evidence against a SN interpretation is the persistently featureless nature of AT2019cmw's spectra, even out to late times. In the case of a supernova, as the photosphere recedes, the radiation would begin to pass through optically thin ejecta and a nebular spectrum would start to be produced. This can be seen in Figure~\ref{fig:Spectra_cmw}, with SN\,2010gx \citep{pastorello_ultra-bright_2010} showing more prominent spectral features as it evolves, whereas AT2019cmw stays featureless. Therefore, it is unlikely that AT2019cmw's properties can be explained using a peculiar SLSN scenario. 

\subsection{Peculiar AGN flare}
\label{sec:AGNrevisit}

As we state in Section~\ref{sec:Characteristics}, Section~\ref{sec:Colours} and Section~\ref{sec:Host}, AT2019cmw's blue UV colour at peak, featureless spectra, as well as the source location's NEOWISE nonvariability and host SED consistent with a normal non-active galaxy \citep{yao_tidal_2023} are evidence against it being the result of an AGN flare. However, due to its nuclear location and high luminosity compared with the majority of previously discovered TDEs and SNe to date, we consider this possibility.

Quasars, for example, have been observed at luminosities significantly higher than AT2019cmw at peak from low to high redshifts \citep{shen_bolometric_2020}. However, as can be seen in the composite spectrum from \citet{vanden_berk_composite_2001} shown in Figure~\ref{fig:Spectra_cmw}, quasars show a combination of broad and narrow emission lines in their spectra. This is unlike our spectral sequence seen in Figure~\ref{fig:Spectra_cmw}, which remains a featureless continuum for its entire evolution. Our spectra are also unlike those of previously observed Seyfert galaxy nuclei, which also show characteristic emission features \citep{mullaney_optical_2008}.

Blazars however have been seen to display predominantly featureless continua in the optical \citep{dupuy_optical_1969,stickel_complete_1991,goldoni_optical_2021}, and are also preferentially seen in massive early type host galaxies \citep{urry_hubble_2000}. However, in the optical their SEDs are dominated by synchrotron emission and appear as a power-law continuum \citep{fiorucci_continuum_2004,ghisellini_transition_2011}, rather than the blackbody SEDs which we observe for AT2019cmw as highlighted in Figure~\ref{fig:SEDs}.

It is possible that AT2019cmw is a peculiar AGN `turn-on' event in a previously quiescent galaxy. \citet{yan_rapid_2019}, for example, observed a prompt AGN turn-on event in the previously quiescent galaxy SDSS1115+0544. However, there are significant observational differences between AT2019cmw and this event. They note the AGN turn-on event in SDSS1115+0544 is unlike previously observed SNe and TDEs due to its strong, persistent UV emission years after the initial `turn-on', indicative of a newly formed accretion disk. This presents as a steep slope into the UV in their photometry and spectroscopy blueward of $\sim4000\,{\AA}$. As seen in Figure~\ref{fig:Spectra_cmw}, we instead observe strong UV flux at early times that fades to show a quiescent galaxy spectrum at late times. Additionally, they observe the formation of prominent hydrogen, oxygen and iron emission lines during the flare, in contrast to AT2019cmw's persistent featureless spectra.

Another notable class of events as a comparison are ANTs, which appear to display smooth lightcurve evolution post-peak \citep{wiseman_systematically_2025}, somewhat similar to AT2019cmw. Although \citet{wiseman_systematically_2025} hypothesise that ANTs are potentially the result of TDEs of intermediate to high mass stars, they appear spectroscopically similar to AGN. \citet{wiseman_systematically_2025} note that the events in their sample ubiquitously show broad AGN-like Balmer emission features. The related subclass of `Extreme Nuclear Transients' (ENTs) presented by \citet{hinkle_most_2025}, which they also hypothesise resulted from TDEs of intermediate to high mass stars, also appear to display broad $\mathrm{Mg\,II}$ emission. \citet{hinkle_most_2025} state that as this feature is commonplace in AGN spectra, this could be the result of a pre-existing reservoir of gas around the SMBH. These events appear distinct from AT2019cmw, which displays featureless spectra throughout its entire observed evolution.

Due to AT2019cmw's observable characteristics being inconsistent with being the result of an AGN-related phenomenon, we consider it inconsistent with being the result of an AGN flare.

\subsection{Tidal Disruption Event}
\label{sec:TDEinterp}

AT2019cmw appears nuclear in its host. Our measured offset of 0.049" and a scatter of 0.112", as detailed in Section~\ref{sec:offset}, is consistent with the offset of $\leq$4" for TDEs selected by the ZTF sample studies from \citet{van_velzen_seventeen_2021} and \citet{hammerstein_final_2023}. As AT2019cmw's host also displays a lack of detected AGN activity, and AT2019cmw itself does not show any AGN-like or SN-like spectral features, we return to the scenario of a TDE to explain its observed properties.

As can be seen in Figure~\ref{fig:BBfits} and Figure~\ref{fig:Colours}, AT2019cmw cools significantly post-peak compared to the featureless TDEs from \citet{hammerstein_final_2023} that display similar featureless spectra. However, as peculiar TDEs and candidate-TDEs have also displayed similar post-peak temperature evolution, we do not consider this to be evidence against a TDE interpretation for AT2019cmw. For example, the candidate featureless TDE Dougie \citep{vinko_luminous_2014} as well as the fast evolving TDE AT2020neh \citep{angus_fast-rising_2022}, both discussed further in Section~\ref{sec:TDEcomparison}, appear to cool significantly post-peak. 

\citet{gezari_tidal_2021} make the argument that the $NUV-r$ colour of a TDE at peak is a potentially robust method of photometrically differentiating TDEs from other transients such as AGN and SNe. AT2019cmw displays significant reddening of its $uvw2-r$ colour post-peak. When measuring colours from interpolated lightcurves of significant detections, from 14.1 and 33.8 days post-peak in its rest-frame the $UVW2-r$ colour reddens from 0.37 mag to 0.73 mag with a consistent absolute magnitude in $UVW2$ of $M < -22$. This would put it in-between the regime of TDEs and AGN. Figure 10 from \citet{gezari_tidal_2021} for instance shows that TDEs appear to have a peak $NUV-r$ colour $\lesssim0$ mag, whereas AGN have a colour of $\gtrsim1$ mag, and SNe a colour of $\gtrsim0.5$ mag. However, as it was rapidly reddening in $u-g$ from at least $\sim6$ days before our first $UVW2$ detection, it is likely that at peak its colour is close to $UVW2-r = 0$ mag.

\begin{figure}
\begin{center}
    \includegraphics[width=1\columnwidth]{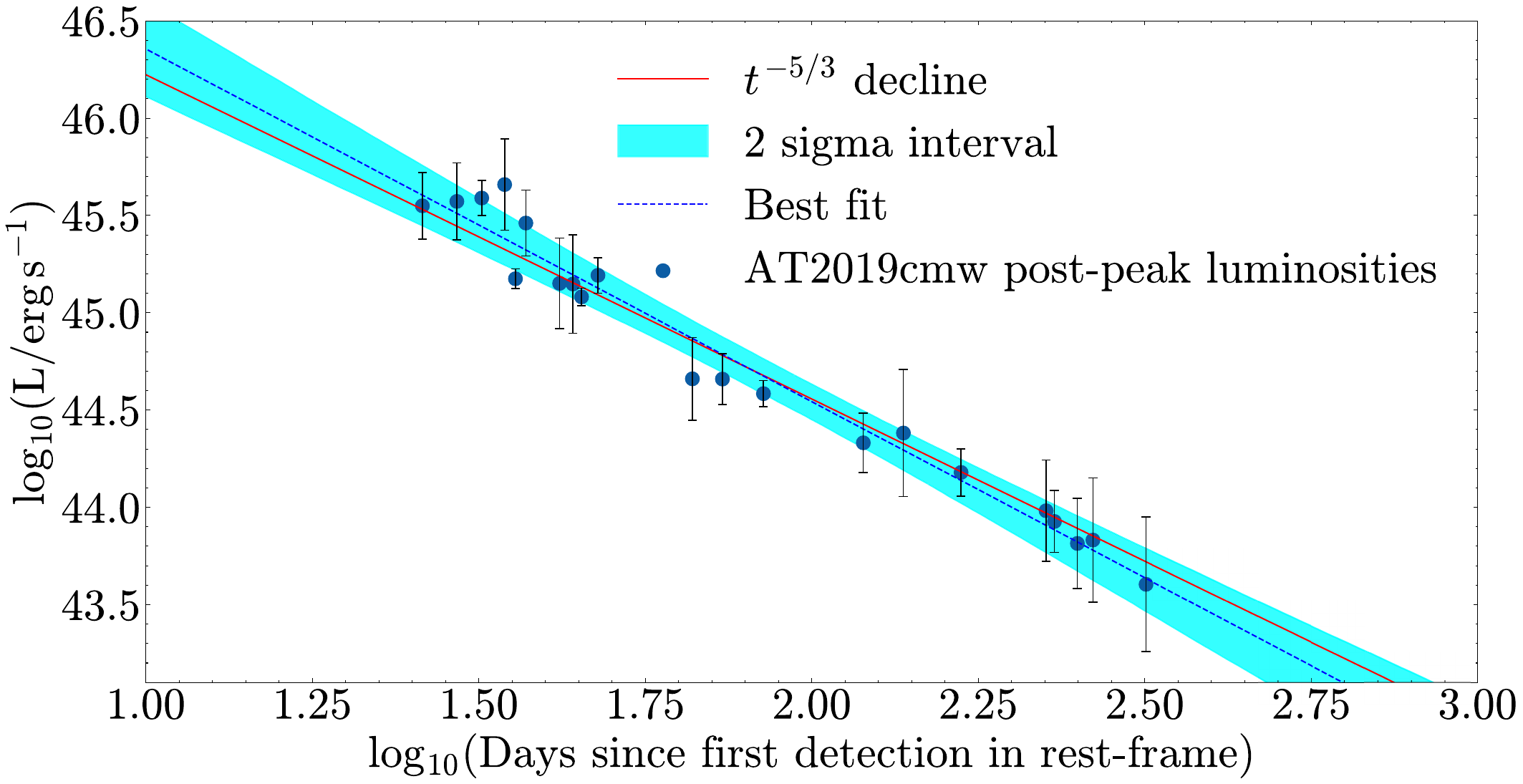}
\end{center}
\caption{\label{fig:PowerLaw} Log-space plot of AT2019cmw's inferred post-peak luminosities from Figure~\ref{fig:BBfits} (blue points) fit to a power-law decline (blue dashed line) with the $2\sigma$ confidence interval on the fit overlaid (blue shaded region). A $t^{-5/3}$ power-law decline has also been plotted for comparison (red line).}
\end{figure}

In Figure~\ref{fig:PowerLaw}, we fit a power-law decline to our derived post-peak bolometric luminosities of AT2019cmw from Section~\ref{sec:Blackbody} to determine the slope index of the decline $\alpha$, with the time of disruption $(t_0)$ set at the time of the first ZTF $g$-band forced photometry detection. At the $1\sigma$ level, we find a value of $\alpha = {1.81^{\textsuperscript{+0.07}}_{\textsubscript{-0.07}}}$. Within $2\sigma$, we find AT2019cmw's rate of decline to be consistent with the $t^{-5/3}$ decline for TDEs predicted using theoretical fallback accretion rates from \citet{rees_tidal_1988} and \citet{phinney_manifestations_1989}. It should be noted that AT2019cmw's first photometric detection is almost certainly after the true $t_0$. Varying $t_0$, we find that AT2019cmw's rate of decline is still consistent with a $t^{-5/3}$ decline within $2\sigma$ if $t_0$ is set to 4 days pre-first detection.

AT2019cmw displays a blue colour and high temperature at peak, and its post-peak luminosity decline rate is consistent with a $t^{-5/3}$ power-law. Combined with its featureless spectra and nuclear location in its host with a lack of any detected AGN activity, this motivates us to interpret AT2019cmw as a peculiar featureless TDE of the class defined by \citet{hammerstein_final_2023}.

\section{Redback `cooling envelope' model}
\label{sec:RedbackIntro}
 
In order to derive parameters such as the mass of the disrupted star and disrupting black hole, we fit our photometry of AT2019cmw using the open-source software package \textsc{Redback} \citep{sarin_redback_2024}. We use the TDE cooling envelope model \citep{metzger_cooling_2022} using the \textsc{Dynesty} sampler \citep{speagle_dynesty_2020} wrapped with \textsc{Bilby} \citep{ashton_bilby_2019}. One advantage of this model compared with other widely used TDE models such as from \citet{mockler_weighing_2019}, is it is not agnostic to the mechanism of energy production from the fallback of stellar material. As described in \citet{sarin_tidal_2024}, it is based on the assumption that the super-Eddington fallback of stellar material onto a black hole during a TDE forms a pressure-supported envelope after the first fallback time, which then cools and undergoes Kelvin-Helmholtz contraction before forming a disk at later times. As such, this model is potentially able to explain how some TDE systems do not brighten in X-rays \citep{yao_tidal_2022} and radio \citep{cendes_ubiquitous_2024} until long after the optical/UV peak, which may apply in the case of AT2019cmw with its X-ray non-detection close to peak and relatively non-constraining radio upper limit at late times.

We adopt the method used by \citet{sarin_tidal_2024} to model their sample of TDEs. The predicted fallback timescale of the system is defined by Equation 3 from \citet{sarin_tidal_2024}, which they adapt from Equation 51 from \citet{stone_consequences_2013}:

\begin{equation}
\label{Fallback_timescale}
\begin{aligned}
{t_{fb} \approx 58\rm{d}{\left(\frac{\kappa}{0.8}\right)^{-3/2}}{m^{1/5}_{*}}M^{1/2}_{bh, 6}},
\end{aligned}
\end{equation}

in which $m_*$ is the mass of the disrupted star in $M_{\odot}$ and $M_{bh, 6}$ is the mass of the disrupting black hole in $10^{6}M_{\odot}$. $\kappa$ is a constant related to the stellar binding energy, which \citet{sarin_tidal_2024} set to 0.8.

Before the envelope formation time in the model as defined by ${\chi}t_{fb}$, with $\chi$ accommodating for model uncertainties in the envelope formation time, we use a phenomenological broken power-law to fit the transient's photometry prior to and shortly after peak. The lightcurve evolution during this time is defined by the Equation 5 from \citet{sarin_tidal_2024}:

\begin{equation}
\label{Broke_power_law}
\begin{aligned}
\begin{aligned}
{F(\nu,t) = A_{\nu}{\left(1-e^{t/t_{peak}}\right)}^{\alpha_{1}}{\left(\frac{t}{t_{peak}}\right)}^{-\alpha_{2}}}.
\end{aligned}
\end{aligned}
\end{equation}

$A_{\nu}$ represents a normalisation for each specific frequency in the model, $\alpha_{1}$ and $\alpha_{2}$ represent the rise and decline indices prior to and post-peak respectively. After $t = {\chi}t_{fb}$, the phenomenological model is smoothly connected to the cooling envelope model.

The uncertainties quoted for our model posterior distributions in this section are $95\%$ credible intervals.

\subsection{Multi-band photometry fit}
\label{sec:Redback}

\begin{figure}
\begin{center}
    \includegraphics[width=1\columnwidth]{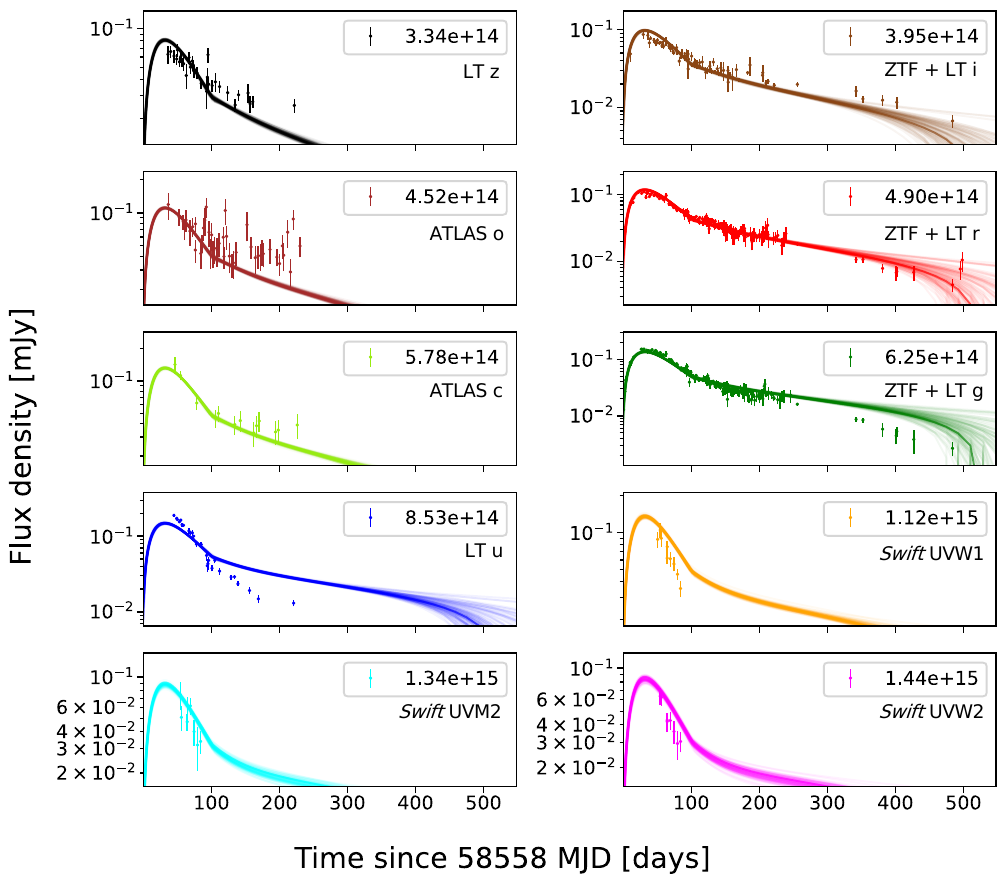}
\end{center}
\caption{\label{fig:RedbackLC} ATLAS, ZTF, LT and \textit{Swift} photometry detailed in Section~\ref{sec:Observations} in ascending order of frequency, overlaid with lightcurves generated from 100 random draws from the posterior. ZTF and LT lightcurves in the same respective bands have been merged. The rest-frame frequency in Hz, as well as the observed photometric filters, for each wavelength bin fit in our model are displayed in the upper-right of each box.}
\end{figure}

We use a broad set of priors for all parameters and run until convergence. Figure~\ref{fig:RedbackLC} shows lightcurves from our model fits plotted against AT2019cmw's photometry. Around peak, the power-law fit is a good match to our data. Although as the model progresses to late times the $g$-band is somewhat overestimated, and $i$-band and $z$-band are underestimated, likely due to the fact that the photospheric cooling seen in AT2019cmw is not predicted to this extent by the model from \citet{metzger_cooling_2022}. The $u$-band and \textit{Swift} UV bands are moderately well fit close to peak, but are significantly overestimated as the model progresses to later times. This is discussed further in Section~\ref{sec:RedbackLimitations}.

\begin{figure*}
\begin{center}
    \includegraphics[width=2\columnwidth]{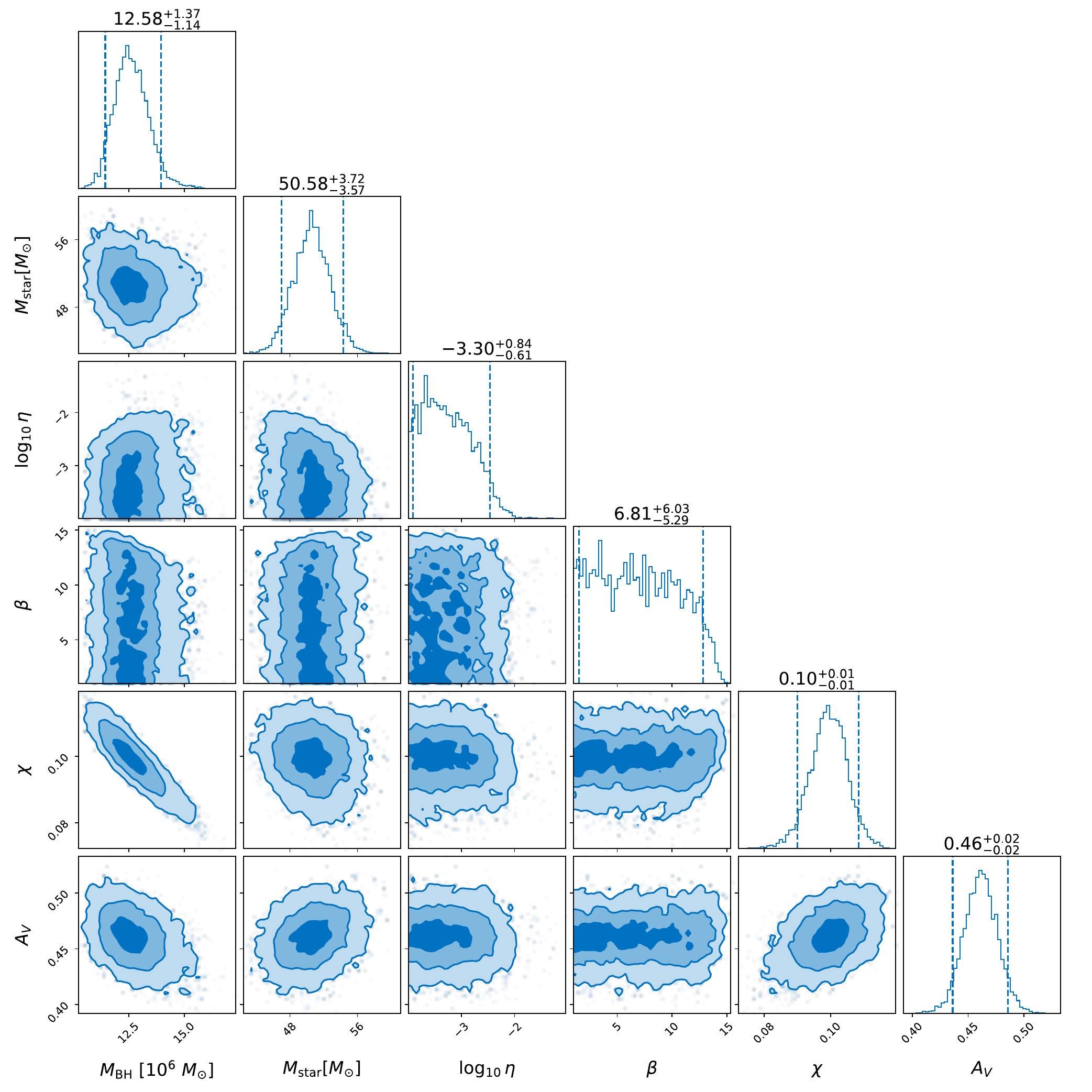}
\end{center}
\caption{\label{fig:Corner} Inferred parameters from our `cooling envelope' model fit to the multi-band photometry shown in Figure~\ref{fig:RedbackLC}. Plot made using \textsc{corner} \citep{foreman-mackey_cornerpy_2016}.}
\end{figure*}

Parameters extracted from this model are displayed in Figure~\ref{fig:Corner} and Table~\ref{tab:AT2019cmw_models_summary}. Our model fits output a disrupted star mass of $M_{*} = {50.58^{\textsuperscript{+3.72}}_{\textsubscript{-3.57}}}\mathrm{M_{\odot}}$, a disrupting black hole mass of $M_{bh} = {{1.258^{\textsuperscript{+0.137}}_{\textsubscript{-0.114}}}\times10^{7}}\mathrm{M_{\odot}}$ and a penetration factor of $\beta = 6.81^{\textsuperscript{+6.03}}_{\textsubscript{-5.29}}$, with $\beta = {r_t}/{r_p}$, the ratio between the radius of the tidal disruption and the periapsis of the in-falling debris respectively. We also find a SMBH feedback efficiency of $\eta = 5^{+30}_{-4}\times10^{-4}$ and an internal extinction of $A_{V} = 0.46^{\textsuperscript{+0.02}}_{\textsubscript{-0.02}}$ mag. We note that these uncertainties are underestimations of their true values, as they do not take into account systematic errors or model misspecification. The complete posterior distribution extracted from this model is presented in Figure~\ref{fig:CornerMultibandFull}.

Although our estimate of $M_{bh}$ is of the same order of magnitude, it is significantly different from the mass of $M_{bh} = {{6.31^{\textsuperscript{+7.82}}_{\textsubscript{-3.50}}}\times10^{7}}\mathrm{M_{\odot}}$ found by \citet{mummery_fundamental_2024} using a late-time thin-disk emission model, and the mass of $M_{bh} = {{8.71^{\textsuperscript{+14.10}}_{\textsubscript{-5.40}}}\times10^{7}}\mathrm{M_{\odot}}$ found by \citet{yao_tidal_2023} using host galaxy scaling relations, with $M_{gal}$ derived from SED fitting of host photometry. However, we note that for both of these values of $M_{bh}$ the uncertainties quoted are at the $1\sigma$ level.

By generating a predictive posterior distribution from 1000 samples of our results, we also derive proxy X-ray luminosities predicted by the model for AT2019cmw. The X-ray luminosities predicted by our model remain on a plateau at $\sim5\times10^{43}$\,erg\,s\textsuperscript{-1} until $\sim100$ days post-envelope formation. They are then predicted to rise and approach the Eddington luminosity of the system as the envelope terminates and the SMBH accretion rate increases \citep{metzger_cooling_2022}, reaching $\sim2\times10^{46}$\,erg\,s\textsuperscript{-1} in our model $\sim1100$ days post-first fallback. These proxy X-ray luminosities are presented in Figure~\ref{fig:ProxyXray}.

Our \textit{Swift} XRT observations, detailed in Section~\ref{sec:Swift}, were taken between $\sim2$ days prior to and $\sim24$ days after ${\chi}t_{fb}$ in the source frame. When comparing the predicted plateau at this phase of evolution, our observed \textit{Swift} X-ray luminosity upper-limit of $\lessapprox2\times10^{43}$\,erg\,s\textsuperscript{-1} is significantly lower than that predicted by our model. However, as this occurs during the `cooling envelope' phase in our model, the predicted X-ray emission would only be detectable from select viewing angles roughly aligned with the black hole's axis of rotation \citep{sarin_tidal_2024} and thus may not have been visible along our line of sight. 

\subsection{Multi-band fit caveats}
\label{sec:RedbackLimitations}

As seen in Figure~\ref{fig:RedbackLC}, a phenomenological power-law fit before and shortly after peak followed by a fit using the cooling envelope model describes the behaviour of our optical and UV lightcurves reasonably well. However, at later times the model underestimates the rate of reddening that we observe, leading us to underestimate the observed rate of cooling from Section~\ref{sec:Blackbody} in our fit. One possible explanation for this is that, as can be seen in Figure~\ref{fig:RedbackLC}, slower fading redder bands have better coverage and S/N at late times. As a result, redder bands could have been more significantly weighted in the fit compared to bluer bands, causing the latter's luminosity to be overestimated at late times. Although the model generally does well at predicting AT2019cmw's photometric evolution, the discrepancy between the colour of our observations and the model fit at later times, as well as the fact that we observe an increase in photometric radius in Section~\ref{sec:Blackbody}, suggest that our value of $\eta$ (the SMBH feedback efficiency) in this fit may be an underestimate.

A significant caveat also applies to our derived value of $M_{bh}$ from Section~\ref{sec:Redback}. As stated in Section 3.1.1 of \citet{metzger_cooling_2022}, close to peak the bolometric luminosity in the cooling envelope can be considered as $L = L_{edd}+L_{fb}$, where $L_{edd}$ is the Eddington luminosity of the envelope and $L_{fb}$ is an additional luminosity contribution from the fallback of stellar material. At later times, the bolometric luminosity in the model approaches $L \approx L_{edd}$, which can be used to constrain the SMBH mass. Randomly sampling the posterior distribution from our fit, we find that the model predicts $L_{edd} \approx 10^{45.3}\rm{erg\,s^{-1}}$ at $\sim300$ days post-envelope formation. In reality, our blackbody fits from Section~\ref{sec:Blackbody} show that the bolometric luminosity decreases to $\sim10^{43.6}\rm{erg\,s^{-1}}$ $\sim300$ days post-peak. This implies that in our fits from Section~\ref{sec:Redback} our value of $M_{bh}$ may be significantly overestimated, as these fits significantly overestimate the bolometric luminosity (and thus $L_{edd}$) at later times. The predicted blackbody characteristics of our model fit from Section~\ref{sec:Redback} are presented in Figure~\ref{fig:MultibandBlackbody}.

As stated in Section~\ref{sec:Redback}, our best fitting model converges to a penetration factor of $\beta = {6.81^{\textsuperscript{+4.34}}_{\textsubscript{-4.01}}}$. Although there is a large uncertainty on this value, it suggests that this event may be a `high-penetration' TDE. Using the following equation for the radius of tidal disruption from \citet{hills_possible_1975}, $r_{t} = {{\left( \frac{M_{bh}}{M_*} \right)}^{1/3}}R_{*}$, the Schwarzschild radius as definied by $R_s = \frac{2GM}{c^2}$ and the approximate mass-radius relation for main sequence stars as defined by $R_* \approx {\left( \frac{M_*}{M_{\odot}} \right)^{4/5}}R_{\odot}$, we find a periapsis, $r_p$, of the stellar material of ${r_{p} = {3.99^{\textsuperscript{+5.72}}_{\textsubscript{-1.55}}}}R_{s}$. Within this range, general relativistic precession could be significant enough to cause infalling debris stream to intersect itself when it completes an orbit \citep{andalman_tidal_2022}. The resulting stream-stream collisions would add an additional source of radiation that is not currently taken into account in the cooling envelope model.

In the same vein, as stated by \citet{metzger_cooling_2022}, $\kappa$ weakly depends on $\beta$ \citep{stone_consequences_2013,guillochon_hydrodynamical_2013}. In the cooling envelope model however, \citet{sarin_tidal_2024} and \citet{metzger_cooling_2022} fix the value of $\kappa$ to 0.8 as an approximation, which corresponds to $\beta=1$ for a polytropic star with $\lambda=5/3$, even though $\beta$ is allowed to vary in the model. Considering Equation~\ref{Fallback_timescale}, this implies that the fallback time may be inaccurate. This is supported by our estimated value of $\chi \sim 0.1$, which implies that the fallback time found using Equation~\ref{Fallback_timescale} may be a significant overestimate. As is discussed in Appendix C of \citet{sarin_tidal_2024}, varying $\kappa$ can also significantly affect estimates of the mass of the star and disrupting black hole. If the true physical value of $\kappa$ is higher, then this could mean that our models estimates of $M_*$ and $M_{bh}$ are underestimated and overestimated respectively. Our results highlight a degeneracy between $\kappa$ and the inferred masses. However, properly accounting for this degeneracy requires modelling how $\kappa$ depends on the star mass and $\beta$, which is not properly included in the Redback implementation of the cooling envelope model. Leaving $\kappa$ as a free parameter will artificially broaden the posterior, but not correctly capture the physics, muddying the physical meaning of some of the implicitly derived quantities in the model. As such, we chose to keep $\kappa$ fixed in our modelling. With a different value of $\kappa$, such as 0.4 as chosen by \citet{sarin_tidal_2024}, we would expect a similar scaling for our derived values of $M_*$ and $M_{bh}$ of $\sim2$ times larger and $\sim4$ times smaller respectively. This would not qualitatively affect our findings from Section~\ref{sec:Redback}.

As seen in Figure~\ref{fig:Corner}, our model fits estimate an internal extinction of $A_{V} = 0.51^{\textsuperscript{+0.02}}_{\textsubscript{-0.02}}$ mag. In Section~\ref{sec:UVSuppression} we describe our observations of a $u$-band excess and a potential UV underluminosity at early times, which we go on to discuss in Section~\ref{sec:UVBehaviour}, as being potentially due to unseen absorption features in the UV. As a result, our estimate of host extinction, and thus the luminosity of the transient in the model, may be an overestimate. Combined with the assumption of $\kappa = 0.8$, it is likely that the disrupted star mass in the model is somewhat overestimated.

\subsection{Bolometric fit}
\label{sec:RedbackBolometric}

\begin{figure}
\begin{center}
    \includegraphics[width=1\columnwidth]{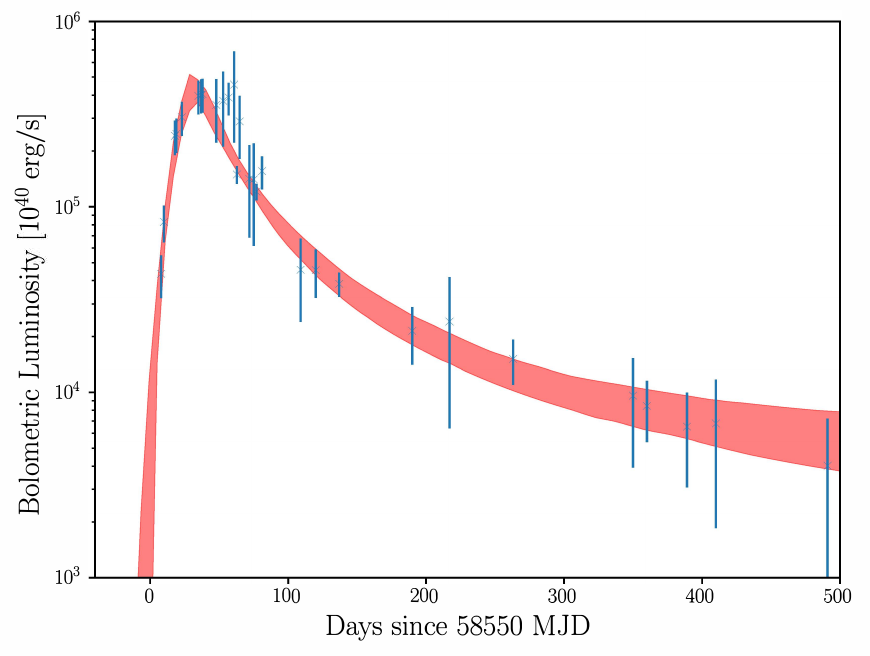}
\end{center}
\caption{\label{fig:RedbackBolometric} $1\sigma$ confidence interval of the cooling envelope model, shown as a red error region, fit to AT2019cmw's derived blackbody luminosities from Section~\protect\ref{sec:Blackbody}, shown as blue error bars.}
\end{figure}

In order to further probe AT2019cmw with the cooling envelope model, we again use the method used in Section~\ref{sec:Redback} to fit our multi-band photometry. However, we instead fit AT2019cmw's derived blackbody luminosities from Section~\ref{sec:Blackbody}. As our input data for this model is achromatic, we cannot constrain the internal extinction as was done in Section~\ref{sec:Redback}, and so assume $A_{V} = 0$ mag. As a result, estimates of other parameters from this model are unaffected by any possible overestimation of this value. The model fit to our derived blackbody luminosities is shown in Figure~\ref{fig:RedbackBolometric}, and key extracted parameters are presented in Table~\ref{tab:AT2019cmw_models_summary}. The complete list of parameters extracted from this model is presented in Figure~\ref{fig:CornerBolometricFull}.

Using this fit, we estimate a disrupted star mass of $M_{*} = {75.30^{\textsuperscript{+21.36}}_{\textsubscript{-36.84}}}\mathrm{M_{\odot}}$. Although our uncertainties are larger here, likely due to us only using epochs with >3 photometric data points as input rather than all of our photometry, this is consistent with the disrupted star mass found by our multi-band fits in Section~\ref{sec:Redback}. As such, this shows that a disrupted star mass in the tens of solar masses is favoured to explain AT2019cmw's characteristics in the context of the cooling envelope model. This would be the case even if the masses of the disrupted star found by our Redback model fits are overestimated by several times compared to the true value (such as if the true value of $\kappa$ is significantly different from 0.8). This is an unprecedentedly high mass for a tidally disrupted star and, if accurate, could have far-reaching implications for our understanding of the environments surrounding SMBHs. This will be discussed further in Section~\ref{sec:StarFormationNucleus}.

For the disrupting black hole in this model we find a mass of $M_{bh} = {{2.4^{\textsuperscript{+2.0}}_{\textsubscript{-1.4}}}\times10^{5}}\mathrm{M_{\odot}}$. This is almost 2 orders of magnitude smaller than the estimate made by our model from Section~\ref{sec:Redback}. As stated in Section~\ref{sec:RedbackLimitations}, this may be due to the assumption in the cooling envelope model that at later times $L \approx L_{edd}$. One possible implication in the context of the cooling envelope model is that at later times, if such an envelope is in fact present, it may be radiating at a rate significantly below $L_{edd}$. This would have the effect of skewing our model's value of $M_{bh}$ to a lower value. A similar discrepancy was also noted by \citet{hajela_eight_2025} when discussing their cooling envelope fit of ASASSN-15oi, which displayed a thermal bolometric luminosity of $L\sim(0.1-0.3)L_{edd}$ at a time when the cooling envelope model predicts $L=L_{edd}$.

Also, curiously, our bolometric luminosity fit is broadly better at predicting the atypical blackbody behaviour that we observe in Section~\ref{sec:Blackbody} than our fits detailed in Section~\ref{sec:Redback}. Despite the lack of colour information in the input data, our model fit generally predicts a photospheric temperature decrease and radius increase/plateau in the first $\sim300$ days post-first fallback. These predicted blackbody characteristics are presented in Figure~\ref{fig:MultibandBlackbody}.

The discrepancy between our cooling envelope model fits will be discussed further in Section~\ref{sec:BHmass}.

\section{Reprocessing-outflow model}
\label{sec:Outflow}

We also consider an alternate reprocessing-outflow model, using the method outlined in \citet{matsumoto_limits_2021}. In contrast to the cooling envelope model discussed in Section~\ref{sec:Redback}, this model operates under the assumption that the emission from optically selected TDEs originates from X-ray photons from the accretion disk reprocessed by an optically thick quasi-spherical expanding outflow. Much like the cooling envelope model discussed in Section~\ref{sec:Redback} this model is motivated by the `low' blackbody temperature observed in optical TDEs, as well as their relatively low X-ray luminosities and late-time radio emission. 

In their model, \citet{matsumoto_limits_2021} determine two radii defined by \citet{nakar_early_2010} and \citet{shen_nature_2015}. The first is the diffusion radius $R_d$, above which radiation can freely escape as the photons diffusion time is shorter than the dynamical time of the ejecta, defined at an optical depth of $\tau{(R_d)} = \frac{c}{v_d}$. The second is the colour radius, at which the photons are last considered to be in thermal equilibrium with the ejecta, defined at an optical depth of ${{\tau}_{\rm{eff}}}(R_c) = 1$.

As input, the model takes the evolution of the transient's luminosity and temperature in order to determine the locations of these radii over time, as well as the outflow density $\rho$ at these radii. These parameters also depend on the outflow velocity $v$, which is varied over a range of possible values as an additional input parameter. As stated by \citet{matsumoto_limits_2021}, for TDEs with outflow velocities lower than the critical velocity $v_c$ at which $R_d = R_c$, $R_c > R_d$. For AT2019cmw this critical velocity lies at $v > 10^{4}\,\mathrm{km\,s^{-1}}$. This is above the approximate upper limit of velocities in the reprocessing-outflow model that \citet{matsumoto_limits_2021} consider, which was set considering observations of TDEs that display broad emission features at a velocity of $v \lesssim 10^{4}\mathrm{km\,s^{-1}} $\citep{arcavi_continuum_2014}. As a result, the rate of mass ejection can be determined for this event using the equation ${\dot{M_{c}}} = 4\pi{{R_c}^2}{\rho_c}{v}$.

\begin{figure}
\begin{center}
    \includegraphics[width=1\columnwidth]{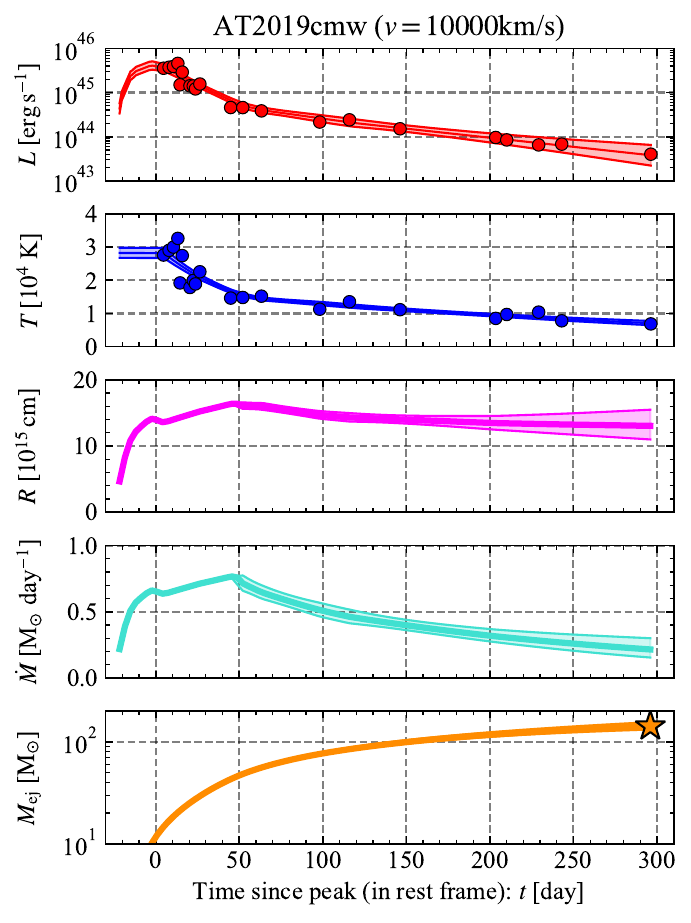}
\end{center}
\caption{\label{fig:Outflow_plots} Input and derived output parameters for our fit of AT2019cmw to the reprocessing-outflow model from $\sim20$ days pre-peak up to $\sim300$ days post-peak. The curves and corresponding error regions plotted show the input luminosities and temperatures used in the model fit, as well as output colour radii, mass-ejection rate per day and cumulative ejecta mass over time. Data points show our derived blackbody characteristics of AT2019cmw from Section~\protect\ref{sec:Blackbody}.}
\end{figure}

\begin{figure}
\begin{center}
    \includegraphics[width=1\columnwidth]{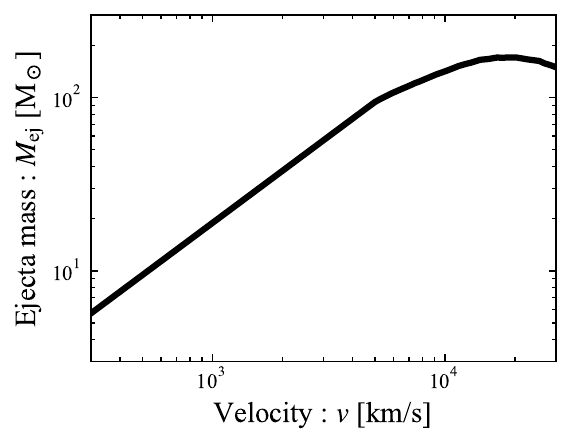}
\end{center}
\caption{\label{fig:Outflow_velocities} Predicted ejecta masses for AT2019cmw in the `reprocessing-outflow' model for a range of assumed input ejecta velocities.}
\end{figure}

We performed an initial fit to our derived blackbody luminosities and temperatures over the first 300 days in AT2019cmw's rest-frame. As can be seen in Figure~\ref{fig:Outflow_plots}, for an outflow velocity of $v = 10^4{\mathrm{km\,s^{-1}}}$ we find an ejecta mass of $\sim142\pm10\,\mathrm{M_{\odot}}$ in the first 300 days post-peak luminosity. Figure~\ref{fig:Outflow_velocities} shows the range of ejecta masses we estimate for a range of input outflow velocities.

Following the method from \citet{matsumoto_limits_2021}, we consider the lower-limit on the outflow velocity in this model to be the escape velocity of the system, for which we use the following relation:

\begin{equation}
\label{Escape_vel_colour}
\begin{aligned}
\begin{aligned}
{v_{esc}}(R_c) \simeq 1900\,\rm{km\,s^{-1}}M^{1/2}_{BH, 6.5}L^{-3/10}_{44}T^{17/20}_{4}.
\end{aligned}
\end{aligned}
\end{equation}

When applying our peak values for temperature and luminosity from Section~\ref{sec:Blackbody}, and the fiducial SMBH mass of $10^{6.5}\,\rm{M_{\odot}}$ used by \citet{matsumoto_limits_2021}, we find an escape velocity at $R_c$ of ${v_{esc}}(R_c) \sim1.5\times10^{3}\,\mathrm{km\,s^{-1}}$. This corresponds to an ejected mass of $\sim30\,\mathrm{M_\odot}$ when applied to Figure~\ref{fig:Outflow_velocities}. Even when applying our lower limit on the SMBH mass from Section~\ref{sec:RedbackBolometric} of $M_{bh} = 1\times10^{5}\,\rm{M_{\odot}}$, we find ${v_{esc}}(R_c) \sim3\times10^{2}\,\mathrm{km\,s^{-1}}$, and thus $\sim6\,\mathrm{M_\odot}$ of ejecta.

A significant caveat of this model is the assumption of quasi-spherical outflows with densities, temperatures and velocities that do not vary with viewing angle. \citet{matsumoto_limits_2021} state that if outflows in TDEs vary strongly with viewing angle, then their observed characteristics could be explained with a lower ejecta mass than calculated for the quasi-spherical case.

Nevertheless, the ejecta mass required to explain AT2019cmw's observed blackbody evolution in this model is $\sim8.5$ times larger than the highest estimate for their sample at an ejecta velocity of $\sim3\times10^{2}\,\mathrm{km\,s^{-1}}$. We note that this is below their approximate lower limit on the outflow velocity of $\sim10^{3}\,\mathrm{km\,s^{-1}}$ for typical SMBHs. Whilst we lack concrete observational evidence of an outflow of the kind described by the reprocessing-outflow model, as we see continuously featureless spectra and a lack of detected late-time radio emission (as will be discussed in Section~\ref{sec:Radiocomparison}), this is evidence that a large amount of material may be needed to explain AT2019cmw's observed characteristics in the context of this model.

This, combined with the high disrupted star masses estimated by our \textsc{Redback} models in Section~\ref{sec:Redback} and Section~\ref{sec:RedbackBolometric}, supports the interpretation that AT2019cmw may be the result of the tidal disruption of a high-mass star.

\begin{table}
\renewcommand{\arraystretch}{1.5}
\caption{Summary of relevant parameters output by our model fits from Section~\ref{sec:Redback}, Section~\ref{sec:RedbackBolometric} and Section~\ref{sec:Outflow}. Uncertainties for parameters from the cooling envelope model are 95\% credible intervals, whereas for the reprocessing outflow model we show $\mathrm{1\sigma}$ errors. A full list of output parameters from our cooling envelope model fits can be seen in Figure~\ref{fig:CornerMultibandFull} and Figure~\ref{fig:CornerBolometricFull}.}
\label{tab:AT2019cmw_models_summary}
\begin{tabular}{lr}
\toprule
\textbf{Cooling envelope model (multi-band fit)} & \\
\midrule
$\mathrm{M_{BH}}$ [$\mathrm{{10^6}M_{\odot}}$] &   $12.58^{{+1.37}}_{{-1.14}}$\\
$\mathrm{M_{*}}$ [$\mathrm{M_{\odot}}$] &   $50.58^{{+3.72}}_{{-3.57}}$\\
$\mathrm{log_{10}\eta}$  &   $-3.30^{{+0.84}}_{{-0.61}}$\\
$\mathrm{\beta}$  &   $6.81^{{+6.03}}_{{-5.29}}$\\
$\mathrm{\chi}$  &   $0.10^{{+0.01}}_{{-0.01}}$\\
$\mathrm{t_0}$ [MJD] &   $58555.63^{{+1.38}}_{{-1.92}}$\\
$\mathrm{A_V}$ [mag] &  $0.46^{{+0.02}}_{{-0.02}}$\\
\toprule
\textbf{Cooling envelope model (bolometric fit)} & \\
\midrule
$\mathrm{M_{BH}}$ [$\mathrm{{10^6}M_{\odot}}$] & $0.24^{{+0.20}}_{{-0.14}}$\\
$\mathrm{M_{*}}$ [$\mathrm{M_{\odot}}$] &   $75.30^{{+21.36}}_{{-36.84}}$\\
$\mathrm{log_{10}\eta}$  &   $-2.02^{{+1.73}}_{{-1.73}}$\\
$\mathrm{\beta}$  &   $51.68^{{+42.51}}_{{-42.91}}$\\
$\mathrm{\chi}$  &   $0.53^{{+0.35}}_{{-0.18}}$\\
$\mathrm{t_0}$ [MJD] &   $58533.68^{{+13.22}}_{{-13.24}}$\\
\toprule
\textbf{Reprocessing-outflow model ($\boldsymbol{\mathrm{10000\,kms^{-1}}}$)} & \\
\midrule
$\mathrm{M_{ej}}$ [$\mathrm{M_{\odot}}$] & $142\pm10$\\
\bottomrule
\end{tabular}
\end{table}

\section{Discussion}
\label{sec:Discussion}

\subsection{Disrupted star mass}
\label{sec:DisruptedStar}

We can infer a conservative minimum value of $\sim0.9\,\mathrm{M_{\odot}}$ for the mass of the disrupted star using the integrated emitted energy post-peak found in Section~\ref{sec:Blackbody}, assuming an accretion efficiency of $10\%$. However, although we note that our mass estimates are dependent on assumptions for values such as $\mathrm{\eta}$, $\mathrm{\kappa}$ and the exact emission geometry of AT2019cmw, our model fits in Section~\ref{sec:Redback}, Section~\ref{sec:RedbackBolometric} and Section~\ref{sec:Outflow} suggest that AT2019cmw is the result of the disruption of a star in the tens of solar masses. In these model fits, this interpretation is favoured to explain AT2019cmw's other observed features, such as its extreme peak luminosity and consistently large blackbody radius. As seen in Equation 1 from \citet{sarin_tidal_2024}, for example, the photospheric radius in the model is proportional to the natural logarithm of the envelope mass.

\begin{figure}
\begin{center}
    \includegraphics[width=1\columnwidth]{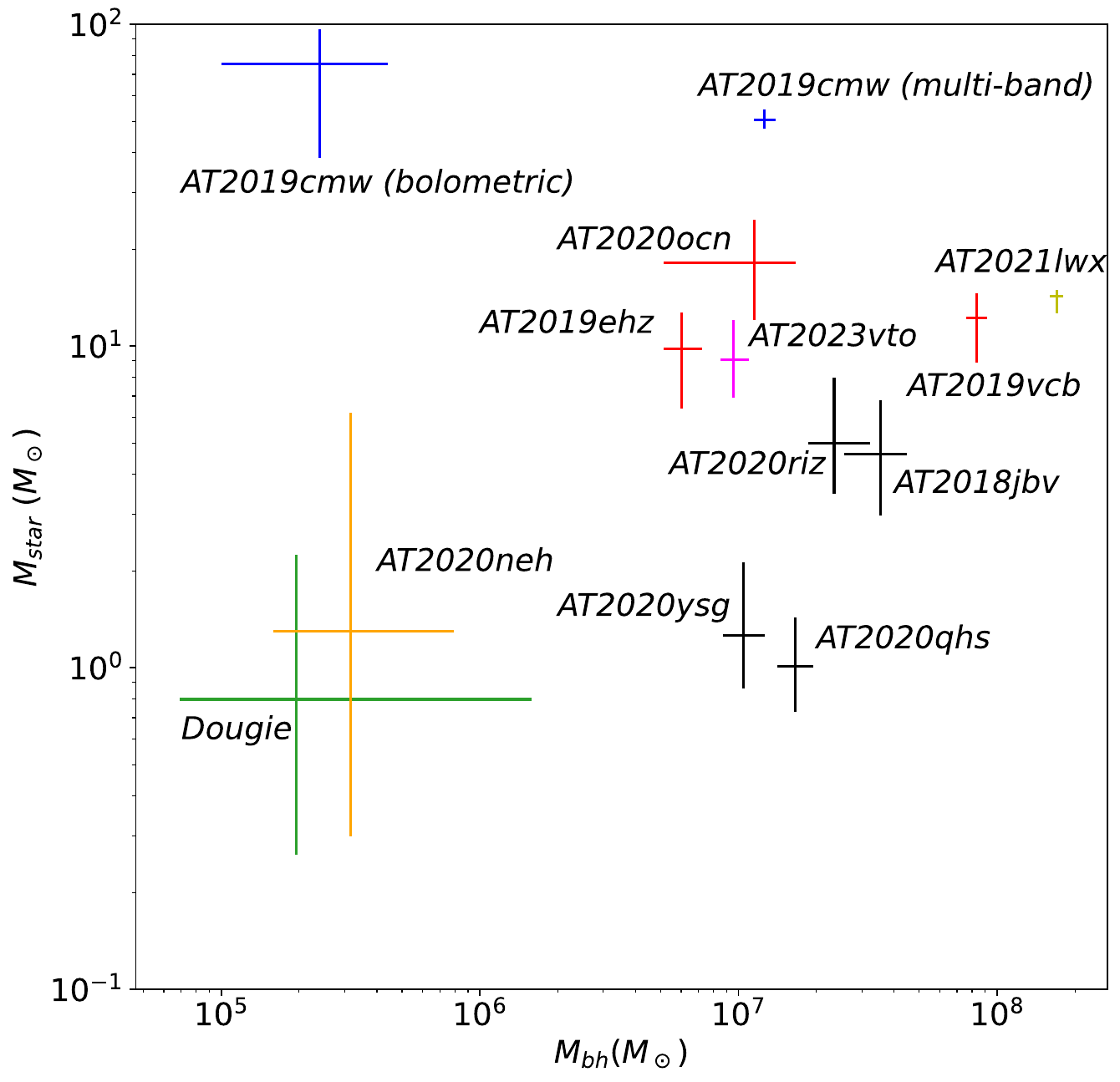}
\end{center}
\caption{\label{fig:BHvsStar} MBH and disrupted star masses from AT2019cmw's Redback model fits and literature TDEs. $95\%$ confidence intervals for AT2019cmw as detailed in Section~\ref{sec:Redback} and Section~\ref{sec:RedbackBolometric} are displayed. Plotted for comparison are AT2021lwx \protect\citep{subrayan_scary_2023}, AT2023vto \protect\citep{kumar_at2023vto_2024}, AT2020neh \protect\citep{angus_fast-rising_2022} and Dougie \protect\citep{vinko_luminous_2014}. The featureless TDEs from \protect\citet{hammerstein_final_2023}, as well as the events with the highest predicted disrupted star masses from their TDE-H (AT2019ehz), TDE-He (AT2020ocn) and TDE-H+He (AT2019vcb) subclass, are also displayed in black and red respectively.}
\end{figure}

Figure~\ref{fig:BHvsStar} shows the values from our model fits for AT2019cmw from Section~\ref{sec:Redback} and Section~\ref{sec:RedbackBolometric} compared to a subsample of TDEs and candidate TDEs with well-measured estimates of the masses of the disrupted star and black hole from the literature. For comparison we chose the featureless TDEs from \citet{hammerstein_final_2023} that appear spectroscopically similar to AT2019cmw, as well as the events with the highest disrupted star masses estimated for each of their other three spectroscopic subclasses. We also chose AT2023vto and AT2021lwx for comparison, recent well-studied events with notably high estimates for the mass of the disrupted star. Additionally, we show AT2020neh \citep{angus_fast-rising_2022} and Dougie \citep{vinko_luminous_2014}, events that appear to show significant post-peak cooling, similar to AT2019cmw. Dougie also displays featureless spectra throughout its evolution.

As can be seen in Figure~\ref{fig:BHvsStar}, the disrupted star masses for AT2019cmw that we infer from our model fits are significantly different from those found by \citet{hammerstein_final_2023} for the featureless TDEs in their sample fit to the model from \citet{mockler_weighing_2019} using \textsc{MOSFiT} \citep{guillochon_mosfit_2018}. The events from \citet{hammerstein_final_2023} all converge to a disrupted star mass of $M_{*} < 10\,\mathrm{M_{\odot}}$ despite being a similar peak luminosity to AT2019cmw. They do however find that TDE-He events converge on average to a high disrupted star mass, with 2 out of 3 of their TDE-He events converging to a stellar mass of $10-20\,\mathrm{M_{\odot}}$ in \textsc{MOSFiT}. They state that this is possibly because they result from the disruption of evolved high mass stars. They also find three TDE H+He events that converge to disrupted star masses of $\gtrsim10\,\mathrm{M_{\odot}}$ in \textsc{MOSFiT}. The event AT2019ehz in their sample also converged to a mass of $\sim10\mathrm{M_{\odot}}$ in \textsc{MOSFiT} despite being a TDE-H event.

Another example is AT2021lwx, which is hypothesised by \citet{subrayan_scary_2023} to be a TDE of a $\sim14\,\mathrm{M_{\odot}}$ star. It was also categorised as an ENT by \citet{hinkle_most_2025}, which they hypothesise are TDEs of $3-10\,\mathrm{M_{\odot}}$ stars. However, as noted by \citet{wiseman_multiwavelength_2023}, in which they explore a non-TDE interpretation of AT2021lwx, the event displays AGN-like emission features. They find that it shows broad Balmer emission, as well as emission features from magnesium, helium and carbon, distinct from events in the TDE subclasses defined by \citet{van_velzen_seventeen_2021}. As we state in Section~\ref{sec:AGNrevisit}, \citet{hinkle_most_2025} hypothesise that the $\mathrm{Mg\,II}$ emission seen in ENTs such as Gaia18cdj, shown in Figure~\ref{fig:Spectra_cmw} for comparison, could be the result of a pre-existing gas reservoir in close proximity to the SMBH. It is possible that the interaction of the stellar debris from a TDE with such a pre-existing gas reservoir could provide an additional source of luminosity, thus increasing the apparent disrupted star mass required to power the flare, but it is currently unknown whether this is the case for ENTs.

In the case of AT2019cmw, it is observationally a featureless event yet still has a high estimated disrupted star mass. Although, any emission lines may be too weak and/or broadened to be detected. Combined with the events discussed above, it is additional evidence that a variety of spectral appearances may result from TDEs of high mass stars, perhaps due to variations in the evolutionary stages of the disrupted stars or environment around the disrupting SMBH. In the case of the TDE-He events from \citet{hammerstein_final_2023} and AT2023vto from \citet{kumar_at2023vto_2024}, which show helium in emission alongside a lack of hydrogen, it is possible that the stars were stripped of their hydrogen envelopes prior to disruption. If this was a result of the stars having ejected their outer layers after leaving the main sequence, as is the case for Wolf-Rayet stars \citep{meynet_stellar_2003}, spectral signatures of heavier elements such as carbon, nitrogen and oxygen may be visible at other wavelengths such as the rest-frame UV. However, as a high mass star would spend a much smaller proportion of its lifetime off of the main sequence than on it, the likelihood of this being the case at the time of disruption is small.

\subsubsection{Implications of a high mass star disruption}
\label{sec:StarFormationNucleus}

The disruption of a star in the tens of solar masses by a SMBH would imply that there is ongoing massive star formation close to the SMBH in AT2019cmw's host, as these stars have lifetimes of only a few million years or less \citep{ekstrom_grids_2012} and so have a very limited time in which to be disrupted by the black hole. AT2019cmw's host appears to be red and poorly star forming, so this raises questions about how the immediate vicinity of its SMBH differs from the surrounding galaxy. If our estimate of the disrupted star mass is correct, it suggests that accreted intergalactic gas, or perhaps an accreted dwarf galaxy, could have caused a small region of massive star formation in the environment close to the SMBH. For instance, the formation of massive stars has been observed at the heart of our own Galaxy, with dozens of young massive star candidates found within the central 2.5 parsecs of the galactic centre by \citet{nishiyama_young_2013}. Smoothed particle hydrodynamic simulations from \citet{bonnell_star_2008} have also showed that star formation from gas infalling into the environment around the SMBH can produce a population with a top-heavy IMF.

Evidence of active massive star formation in early-type galaxies may have been seen in the case of SN2006gy \citep{ofek_sn_2007,smith_sn_2007}. The recent study from \citet{jerkstrand_type_2020} for instance suggests that a common envelope ejection from the merger of a white dwarf and a massive hydrogen-rich star shortly before a Type Ia supernova could explain its extreme luminosity. Centrally concentrated star formation has also been observed in NGC 3182 \citep{pak_origin_2023}, as well as a centrally concentrated molecular outflow in NGC 1266 \citep{alatalo_discovery_2011}, both of which are nearby early-type lenticular galaxies. However, it should be noted that for NGC 3182 and NGC 1266 these phenomena are associated with AGN-driven outflows \citep{alatalo_suppression_2015,pak_origin_2023}, whereas we do not see evidence of an AGN in AT2019cmw's host.

\subsection{Black hole mass}
\label{sec:BHmass}

As can be seen in Figure~\ref{fig:BHvsStar}, we find values of $M_{bh}$ separated by almost two orders of magnitude from our \textsc{Redback} cooling envelope model fits from Section~\ref{sec:Redback} and Section~\ref{sec:RedbackBolometric}, at $M_{bh}\approx1.3\times10^{7}\mathrm{M_{\odot}}$ and $M_{bh}\approx2.4\times10^{5}\mathrm{M_{\odot}}$ respectively. As we discuss in Section~\ref{sec:RedbackBolometric}, combined with the discrepancy between our estimates of $M_{bh}$ and the independent high values of $M_{bh}$ found by \citet{yao_tidal_2023} and \citet{mummery_fundamental_2024} for AT2019cmw, this suggests that at later times the cooling envelope may not accurately describe AT2019cmw's evolution. Although it appears that a high disrupted star mass may be required to explain our near-peak observations, as it evolves AT2019cmw may deviate significantly from the physical assumptions in current TDE models. 

Expanding our comparison to other similar events, such as the featureless TDEs found by \citet{hammerstein_final_2023}, \citep{yao_tidal_2022} as well as the events in the sample by \citet{yao_tidal_2023} \citep{hammerstein_ztf_2021,chu_ztf_2022}, we find that these events are generally predicted to have high black hole masses. \citet{hammerstein_final_2023} note that their host galaxies are significantly larger than those of other TDE subclasses in their sample, with their \textsc{MOSFiT} fits outputting significantly larger black hole masses. However, some events go against this trend. For example, the candidate featureless TDE Dougie \citep{vinko_luminous_2014} appears to have a disrupting black hole mass of $M_{bh} = {{1.95^{\textsuperscript{+13.90}}_{\textsubscript{-1.26}}}\times10^{5}}\mathrm{M_{\odot}}$ from \textsc{TDEFit} modelling \citep{guillochon_ps1-10jh_2014}, which is potential evidence against this being the case for all of these events. A high black hole mass also doesn't appear to always result in a featureless event. As can be seen in Figure~\ref{fig:BHvsStar}, the TDE-H+He AT2019vcb from \citet{hammerstein_final_2023} has a higher value of $M_{bh}$ from \textsc{MOSFiT} than all featureless TDEs in their sample. AT2021lwx, which \citet{subrayan_scary_2023} find is a TDE from a black hole with $M_{bh} = 1.7\pm0.1\times10^8\rm{M_{\odot}}$, and hypothesised by \citet{hinkle_most_2025} to be a TDE around a high mass $(>10^{8}\mathrm{M_{\odot}})$ black hole, has a higher black hole mass than all events from \citet{hammerstein_final_2023}. Due to the diversity of derived black hole masses found for TDEs of various spectral classes, whether there is a physical link between the mass of the disrupting black hole and the spectral appearance of the resulting TDE is currently uncertain.

One possible reason for the diversity of derived black hole masses is systematic model differences. A comparison between values of $M_{bh}$ for 5 TDEs derived using \textsc{MOSFiT} by \citet{hammerstein_final_2023}, \textsc{TDEmass} by \citet{ryu_measuring_2020} and the cooling envelope model in \textsc{Redback} is presented in Table 4 of \citet{sarin_tidal_2024}. For 4 of these events, the predicted values are consistent to within an order of magnitude. However, for the faint and fast event AT2020wey \citet{arcavi_transient_2020}, \citet{charalampopoulos_at_2023} and \citet{hammerstein_final_2023} find a value of $M_{bh}$ approximately 2 orders of magnitude greater than \citet{ryu_measuring_2020} and \citet{sarin_tidal_2024}.

Although AT2019cmw was found to have a high black hole mass by \citet{yao_tidal_2023} and \citet{mummery_fundamental_2024}, the discrepancy between the values of $M_{bh}$ we find from our cooling envelope model fits means that we cannot confirm whether it actually was the result of a TDE from a high mass SMBH. The case of AT2020wey presented by \citet{sarin_tidal_2024} suggests that for events at the extreme ends of the TDE luminosity distribution, such as AT2019cmw, our current models may not be able to accurately measure their black hole masses, and that further theoretical work is needed.

\subsection{Photometric comparison to other TDEs}
\label{sec:TDEcomparison}

\begin{figure}
\begin{center}
    \includegraphics[width=1\columnwidth]{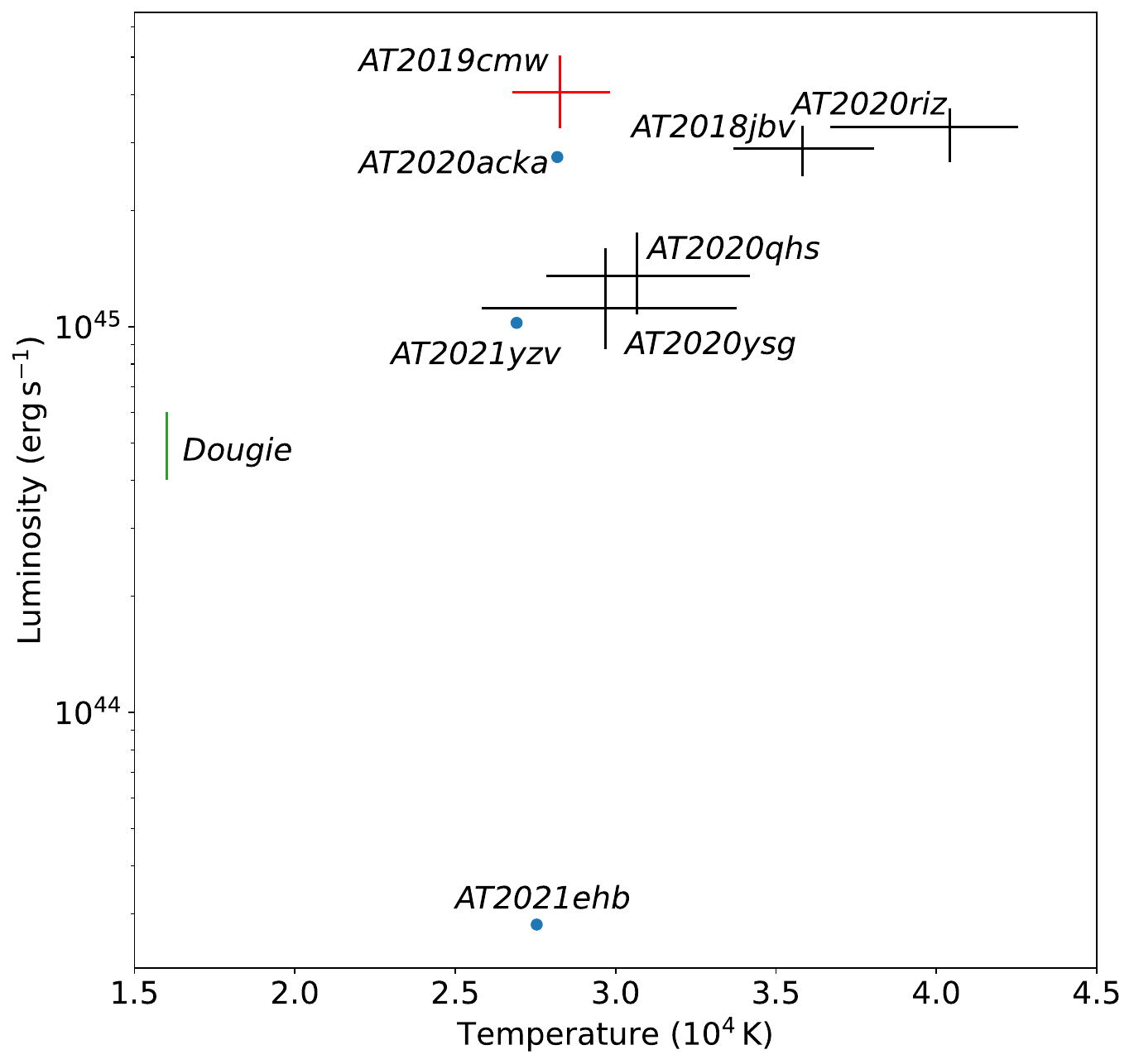}
\end{center}
\caption{\label{fig:tempvslum} Temperatures and luminosities at peak of AT2019cmw and featureless TDEs from the literature. The featureless TDEs from \protect\citet{hammerstein_final_2023} are shown in black, and the 3 featureless events from \citet{yao_tidal_2023} are shown in red. The latter 3 events did not have associated uncertainties for their peak temperatures and luminosities. Dougie \protect\citep{vinko_luminous_2014}, shown in green, also did not have an uncertainty listed for its peak temperature.}
\end{figure}

In Figure~\ref{fig:BHvsStar}, we compare AT2019cmw to AT2021lwx \citep{subrayan_scary_2023,hinkle_most_2025} and AT2023vto \citep{kumar_at2023vto_2024}, two other well-studied TDEs with high disrupted star mass estimates. AT2019cmw's peak luminosity of $\mathrm{log}(L/\mathrm{ergs^{-1}}) \sim 45.6$ that we measure in Section~\ref{sec:Blackbody} is comparable to the luminosity of $\mathrm{log}(L/\mathrm{ergs^{-1}}) = 45.7$ for AT2021lwx \citep{subrayan_scary_2023}, and is significantly higher than AT2023vto's value of $\mathrm{log}(L/\mathrm{ergs^{-1}}) ={44.89^{\textsuperscript{+0.02}}_{\textsubscript{-0.03}}}$ as measured by \citet{kumar_at2023vto_2024}. AT2019cmw's peak temperature of $\mathrm{T} = 10^{4.45}$K is higher than the temperature of $\mathrm{T} = 10^{4.30}$K measured by \citet{kumar_at2023vto_2024} for AT2023vto using spectral blackbody fitting, however they note that this is a conservative lower limit. AT2021lwx's peak temperature of $\mathrm{T} \approx 10^{4.20}$K is somewhat lower than both. In terms of integrated radiated energy over their observed evolution, AT2019cmw and AT2023vto emit similar energies of $\sim1.7\times10^{52}\,\mathrm{erg}$ and $\sim1.8\times10^{52}\,\mathrm{erg}$ respectively. AT2021lwx was measured by \citet{subrayan_scary_2023} to have radiated $\sim9.7\times10^{52}\,\mathrm{erg}$, a much higher total radiated energy than AT2019cmw. Although this could suggest a higher disrupted star mass, it could also be indicative of a higher accretion efficiency. It could also be due to shocks from collisions between stellar debris and pre-existing material around the SMBH, as was already discussed in Section~\ref{sec:DisruptedStar}.

In Figure~\ref{fig:BHvsStar}, we also show AT2019ehz, AT2020ocn and AT2019vcb, the events from the TDE-H, TDE-He and TDE-H+He subclasses respectively with the highest estimated disrupted star masses in the sample from \citet{hammerstein_final_2023}. Notably, even though these events have higher estimates for the mass of the disrupted star than the featureless TDEs (excluding AT2019cmw) they are less luminous at peak, with peak bolometric luminosities between $\mathrm{log}(L/\mathrm{ergs^{-1}}) \sim 43.6$ and $\mathrm{log}(L/\mathrm{ergs^{-1}}) \sim 44.0$. This is potential evidence that that a high disrupted star mass is not what is primarily driving featureless TDEs such as AT2019cmw to display such high luminosities, with other factors such as higher accretion efficiency possibly being more important. AT2019ehz and AT2019vcb are also cooler at peak than AT2019cmw and the featureless TDEs from \citet{hammerstein_final_2023}, at $\sim10^{4.3}$K and $\sim10^{4.4}$K respectively. However AT2020ocn has a comparable peak temperature to the featureless TDEs at $\sim10^{4.5}$K. The case of AT2020ocn shows that featureless spectra may not be a temperature dependent phenomenon.

Whilst AT2019cmw has a peak temperature and luminosity similar to the featureless TDEs found by \citet{hammerstein_final_2023}, as can be seen in Figure~\ref{fig:BBfits}, it seems to have evolved differently after peak. Its prompt cooling in the first 300 days post-peak is in contrast to their lack thereof, as they appear to stay at a constant temperature between $\sim10^{4.4}$K and $\sim10^{4.6}$K. The evolution of AT2019cmw's inferred radius is also unlike other TDEs in this class, as they all display a decrease in their radius over time whereas it does not. In the case of its fading, it appears to be primarily driven by cooling rather than a decrease in its radius as previously seen in most other TDEs.

As seen in Figure~\ref{fig:tempvslum}, expanding our comparison with other featureless TDEs paints a broader picture. The sample from \citet{yao_tidal_2023} contains three additional featureless TDEs not included in \citet{hammerstein_final_2023}'s sample. Of these, AT2020acka \citep{hammerstein_ztf_2021} and AT2021yzv \citep{chu_ztf_2022}, have comparable peak luminosities to those just discussed at $\mathrm{log}(L/\mathrm{ergs^{-1}}) = 45.44$ and $\mathrm{log}(L/\mathrm{ergs^{-1}}) = 45.01$ respectively. AT2021ehb \citep{yao_tidal_2022} however has a peak luminosity of $\mathrm{log}(L/\mathrm{ergs^{-1}}) = 43.54$. As stated by \citet{yao_tidal_2022}, this is evidence that featureless TDEs are not exclusively overluminous, such as in \citet{hammerstein_final_2023}'s sample. Although AT2019cmw is likely another example of an extremely luminous featureless TDE, the rarity of this subclass means that there is a selection bias towards finding more luminous events at high redshift.

Other events also hint at additional possible photometric diversity in this subclass. \citet{yao_tidal_2023} for example finds TDEs from a wide variety of spectral classes with post-peak temperature change, with 11 TDEs in their sample showing significant $uvw2-r$ colour evolution. Another example is AT2020neh \citep{angus_fast-rising_2022}, a fast-evolving TDE with broad hydrogen emission that displays significant post-peak temperature evolution. These events suggest that AT2019cmw's featureless spectral appearance and its atypical blackbody temperature evolution are likely two distinct and unrelated phenomena.

Dougie \citep{vinko_luminous_2014} is an example of an unclassified event that shows similarities to AT2019cmw and the featureless TDEs discussed above. It displayed a featureless continuum throughout its entire visible evolution, with only narrow host features visible at late times. Additionally, Dougie also cooled from $\sim\mathrm{13\,kK}$ to $\sim\mathrm{6.3\,kK}$ from 10 days to 36 days post-peak, similar to the cooling seen in AT2019cmw and AT2020neh \citep{angus_fast-rising_2022}. However, it was significantly less luminous at peak than AT2019cmw and the featureless TDEs from \citet{hammerstein_final_2023}, with $L_{peak} = 5\pm1 \times 10^{44}$erg\,s\textsuperscript{-1}. It was also somewhat cooler than other featureless TDEs at peak at $\approx16\,\rm{kK}$. Notably, \citet{vinko_luminous_2014} found that Dougie was offset from its host nucleus by $\sim1.3^"$, which they state corresponds to $\sim3.9\,\rm{kpc}$. This is in contrast to the nuclear locations of most other TDEs in the literature.

\subsubsection{Luminous fast coolers as TDEs?}
\label{sec:LFCs}

Dougie \citep{vinko_luminous_2014} has been speculated to be part of a possible new class of rapidly evolving transients, termed `luminous fast coolers' (LFCs) by \citet{nicholl_at_2023}, with AT2022aedm and AT2020bot \citep{ho_search_2023} also presented as possible members of this class. These events appear to display many similarities to AT2019cmw. AT2022aedm for instance reaches a peak luminosity of $\sim10^{45}\,\rm{erg\,s^{-1}}$ and displays spectra lacking broad emission features throughout its observed evolution. Notably, it cools post-peak from $\sim30\,\rm{kK}$ to $\sim4\,\rm{kK}$ whilst appearing to show an expanding rather than contracting photosphere, and displays no significantly detected X-ray or radio emission, with Dougie also being undetected in X-ray \citep{vinko_luminous_2014}. All three events also occurred in hosts with low SFRs.

However, as stated by \citet{nicholl_at_2023}, these events are all significantly offset from their host galaxies' nuclei. \citet{nicholl_at_2023} also show that AT2022aedm's spectra are not entirely featureless, displaying narrow emission features at early times and narrow absorption at late times, with \citet{ho_search_2023} also finding that AT2020bot displays unidentified broad spectral features. 

Although some characteristics of LFCs are in contrast to what we find for AT2019cmw, their similarities are still notable. \citet{nicholl_at_2023} speculate that LFCs may be caused by TDEs from intermediate or stellar-mass black holes due to their fast evolution and non-nuclear locations in their hosts. As a result, AT2019cmw may be related to this class of events, possibly representing a `missing link' between LFCs and previously classified TDEs. This, and the colour-changing TDEs discussed in Section~\ref{sec:TDEcomparison}, suggest that there may be extensive photometric in addition to spectral heterogeneity in TDEs. However, the physical mechanism that causes some TDEs to significantly redden in contrast to most other classified events is currently unknown. 

Due to their post-peak colour evolution and often ambiguous spectra, it is possible that other featureless and/or cooling AT2019cmw-like TDEs/LFCs may have gone unclassified, or were misclassified as other classes of transients. AT2020bot, for example, was noted to have unidentified broad spectral features similar to Type Ic-BL SNe by \citet{ho_search_2023}. A larger sample of well-studied events is required to better determine the observational similarities and differences between these subclasses. The ability to accurately distinguish between them photometrically will be essential to uncovering their driving mechanisms, especially with the large number of expected TDE discoveries from the upcoming Vera C. Rubin observatory \citep{ivezic_lsst_2019,bricman_prospects_2020}.

\subsection{Deviations from simple blackbody evolution in the UV}
\label{sec:UVBehaviour}

As described in Section~\ref{sec:Colours}, AT2019cmw's $u-g$ colour increases at a high rate shortly after peak before slowing. For a simple cooling blackbody, one would expect the rate of reddening in $u-g$ to increase rather than decrease as the peak wavelength of the blackbody emission increases past the $u$-band, and the $u$-band increasingly samples the Wien tail of a cooling black body. One possibility is that we are observing emission from inhomogeneous regions of material surrounding the black hole with different temperatures, rather than one homogeneous region with a single temperature. As we state in Section~\ref{sec:UVSuppression}, this possibility motivated us to attempt a multi-component blackbody fit to explain the potential underluminosity we see in the \textit{Swift} UV bands compared with the optical. This did not result in an improvement in our fits.

\citet{vinko_luminous_2014} also observed a UV underluminosity in Dougie, similar to what we observe for AT2019cmw in Section~\ref{sec:UVSuppression}. One possible explanation that they suggested was the appearance of UV absorption features as seen in the case of PS1-11af by \citet{chornock_ultraviolet-bright_2013}. As seen in Figure~\ref{fig:Spectra_cmw}, the spectra that we obtained from LRIS cover a large portion of the same wavelengths as the $u$-band filter on the LT ($1950\,\AA$ to $2632\,\AA$ in rest-frame), and do not show any emerging absorption features. As such, this likely does not explain the rapid $u$-band fading that we observe from AT2019cmw. The physical origin of this behaviour is therefore currently unknown.

However, the presence of UV absorption features at wavelengths not covered by our optical spectroscopy could potentially explain an underluminosity in the \textit{Swift} UV bands that we find hints of in Figure~\ref{fig:SEDs}. As can be seen in Figure 9 of \citet{hung_discovery_2019}, several TDEs display prominent UV spectral features, both in absorption and emission between $\sim1000\,\AA - 2000\,\AA$. A notable example that they show is the TDE iPTF15af which, although its optical spectra presented with hydrogen and helium emission features superimposed on a blue continuum in contrast to AT2019cmw's featureless spectra, showed broad UV absorption troughs from elements such as oxygen and carbon similar to broad absorption line quasars \citep{blagorodnova_broad_2019}. Comprehensive UV spectral follow-up of a larger sample of TDEs would be useful in determining how common such UV spectral features are in TDEs.

Potential UV absorption features are not taken into account in the model fits discussed in Section~\ref{sec:Redback} and Section~\ref{sec:Outflow}, although our \textsc{Redback} model fit suggested significant host extinction/reddening. The presence of any UV absorption features could mean that our value for host extinction, and thus our estimate for the luminosity and disrupted star mass, are overestimated. As stated in Section~\ref{sec:RedbackLimitations} however, when we performed a fit to the Cooling Envelope model using our derived blackbody luminosities from Section~\ref{sec:Blackbody} that neglect any internal extinction or UV suppression, we still predict a high disrupted star mass. As our fit to the reprocessing-outflow model in Section~\ref{sec:Outflow} was also performed using these same luminosities, we can conclude that any UV suppression that may be present is not significantly affecting our estimates of the disrupted star mass.

\subsection{Observational evidence of jets and outflows}
\label{sec:Radiocomparison}

Several TDEs have been observed at extreme luminosities in the radio. Swift J164449.3+573451 \citep{cendes_radio_2021} for example appears to show bright emission from on-axis jets, with its radio luminosity exceeding $10^{40}$erg\,s\textsuperscript{-1} for over 1000 days post-first detection \citep{cendes_ubiquitous_2024}, $\sim2$ order of magnitude greater than our upper limit of ${\nu}{L}_{\nu} = 2.8\times10^{38}$erg\,s\textsuperscript{-1} for AT2019cmw from Section~\ref{sec:XrayResults}. Another event that \citet{cendes_ubiquitous_2024} discuss is AT2022cmc \citep{andreoni_very_2022}, which has a radio luminosity evolution so far very similar to Swift J164449.3+573451 and is thought to also possess an on-axis jet. Optically, AT2022cmc is a useful comparison event in this context, as \citet{andreoni_very_2022} and \citet{hammerstein_jetted_2025} note that after its initial fast-fading red lightcurve it displays a blue, optically luminous plateau with featureless spectra. They suggest a possible link between TDEs with relativistic jets and the featureless TDEs from the sample by \citet{hammerstein_final_2023}. Just as we do not see evidence in the radio, we do not see evidence of jetted optical emission for AT2019cmw as was the case for AT2022cmc. However, this only provides evidence against the presence of on-axis jetted emission in AT2019cmw, as an off-axis jet at a large viewing angle would not display the same observational characteristics as shown by AT2022cmc.

\citet{andreoni_very_2022} state in their study of AT2022cmc that when correcting for beaming effects, the rate that they calculate for on-axis relativistic TDEs means that $\sim1\%$ of TDEs produce relativistic jets. However, \citet{cendes_ubiquitous_2024} observed 23 optically selected TDEs, and detected 10 with radio emission (that did not come from star formation in the host or a pre-existing AGN) with luminosities ranging from $10^{37}$ - $10^{39}$\,erg\,s\textsuperscript{-1}. They state that their observations are at odds with off-axis relativistic jets being able to explain this population of TDEs, and suggest that the emission instead may be explained by a non-relativistic outflow from processes such as late-time disk formation. As can be seen in Figure 1 of \citet{cendes_ubiquitous_2024}, some events in their sample are detected at luminosities an order of magnitude fainter than our upper limit for AT2019cmw at similar times. As such, our upper limit is relatively non-constraining in this context. However, some events in their sample such as ASASSN-14ae \citep{prieto_asas-sn_2014,holoien_asassn-14ae_2014} and AT2018hyz \citep{arcavi_transient_2018,cendes_mildly_2022} were detected at much later times than our observations, with their luminosities continuing to rise over time. ASASSN-14ae for instance was first detected in radio 2313 days post-optical discovery by \citet{cendes_ubiquitous_2024}, with its luminosity increasing to $\sim10^{38}$\,erg\,s\textsuperscript{-1} 930 days later, which is approaching our upper limit for AT2019cmw. 

Although we place tight constraints on the presence of an on-axis jetted component for AT2019cmw, one that is significantly off-axis cannot be ruled out. We also cannot discount that there was radio emission present from a non-relativistic late-launched outflow below our detection limit, or that it may become significantly detectable at later times. As such, continued late-time radio monitoring of AT2019cmw and other similar events is warranted.

\section{Conclusions}
\label{sec:Conclusions}

We have presented a comprehensive analysis of the extraordinary spectroscopically featureless nuclear transient AT2019cmw. Following an intense photometric and spectroscopic follow-up campaign, we found it to have a peak luminosity of $\rm{L_{peak}} \sim 10^{45.6}$\,erg\,s$^{-1}$ from blackbody SED fitting, making it one of the most luminous thermal transients ever discovered. Its post-peak photometric evolution is peculiar when compared with most other TDEs. Additionally, our model fits to its photometry and luminosity evolution imply the involvement of a massive star in close proximity to its host's SMBH.

\begin{itemize}
\item Although AT2019cmw's relatively fast rise, extreme luminosity and post-peak cooling could theoretically be explained with the scenario of a highly superluminous supernova, the combination of its lack of visible spectral features throughout its evolution as well as its blackbody temperature and radius evolution make it difficult to explain in this context. The host galaxy's SED from \citet{yao_tidal_2023}, the lack of any detected AGN spectral lines and the absence of significant detected activity attributable to an AGN in forced photometry also suggests that it is unlikely that there is a preexisting AGN in the host. This, alongside its blue colour at peak and $t^{-5/3}$ power-law luminosity decline post-peak, is consistent with AT2019cmw being a peculiar featureless TDE.

\item Our cooling envelope model fits in Section~\ref{sec:Redback} and Section~\ref{sec:RedbackBolometric} converge to SMBH masses that deviate by $\sim2$ orders of magnitude. As we state in Section~\ref{sec:BHmass}, this implies that the physical assumptions in current TDE models are not sufficient to explain the photometric evolution of AT2019cmw, especially at later times. This is also highlighted by our difficulty in reproducing AT2019cmw's post-peak colour evolution in our fit of its multi-band photometry in Section~\ref{sec:Redback}. Alongside the previously noted systematic differences between parameters estimated by different TDE models for the extreme event AT2020wey \citep{sarin_tidal_2024}, this suggests that additional theoretical work is needed to improve current TDE models.

\item We determine a conservative lower-limit of $\sim0.9\,\mathrm{M_{\odot}}$ for the mass of the disrupted star from a measured radiated energy of $\sim1.7\times10^{52}\,\mathrm{erg}$. However, our photometric fits using both the `cooling envelope' and `reprocessing-outflow' models suggest that AT2019cmw was the result of the disruption of a $>10\,\mathrm{M_{\odot}}$ star. Although our models may be somewhat overestimating the mass of the disrupted star when compared to reality its physical characteristics, such as its consistently large blackbody radius, suggest that a high mass star is likely required to explain our observations of AT2019cmw. A TDE of a high mass star suggests that although AT2019cmw's host is broadly poorly star forming, it may possess a region of active star formation close to its SMBH, potentially due to accreted intergalactic gas. Our high estimate for AT2019cmw's disrupted star mass, in addition to the potential TDEs of high mass stars from \citet{hammerstein_final_2023}, \citet{hinkle_most_2025} and \citet{wiseman_systematically_2025}, could point to a top-heavy IMF in close proximity to some SMBHs, as suggested by the simulations from \citet{bonnell_star_2008}.

\item AT2019cmw's SED shows a transient near-UV underluminosity shortly post-peak, as was seen in the candidate-TDE Dougie \citep{vinko_luminous_2014}. It also appears to show $u-g$ reddening at a faster rate than is expected for a simple cooling blackbody. The physical origins of these behaviours are currently unknown. This points to the need for comprehensive UV photometric follow-up of similar events, as well as UV spectroscopic follow-up, in order to probe for deviations from a simple blackbody SED and/or the presence of absorption features in the UV.

\item Our radio upper-limit of ${\nu}{L}_{\nu} < 2.8\times10^{38}$erg\,s\textsuperscript{-1} at 971 days post-peak in rest frame, as well as the lack of an initial red, fast-fading phase of its optical lightcurve as was seen in the case of AT2022cmc \citep{andreoni_very_2022}, places tight constraints on the presence of any early on-axis relativistic jetted emission from AT2019cmw. However, observations from \citet{cendes_ubiquitous_2024} suggests that some TDEs can first display rising radio emission thousands of days post-first optical detection, possibly from non-relativistic outflows. This points to the need for continuing deep radio follow-up of AT2019cmw and other similar events in order to search for evidence of late-launched outflows that may become visible at later times.

\item Our \textit{Swift} XRT X-ray upper limit of $\gtrsim3\times10^{43}$\,erg\,s\textsuperscript{-1}, observed between $\sim12$ and $\sim54$ days post-peak in AT2019cmw's rest-frame, is not particularly constraining on the presence of X-ray emission. However, the ratio between AT2019cmw's derived blackbody luminosity and its X-ray luminosity ($L_{BB}/L_X$) upper limit was $\sim50-100$. In comparison, 6 out of 8 of the X-ray detected TDEs observed in the sample from \citet{hammerstein_final_2023} at a similar time post-peak showed $L_{BB}/L_X\lesssim50$. This implies that AT2019cmw was not an X-ray luminous event at the time of our observations.

\item Alongside other similar events, its atypical post-peak blackbody temperature and radius evolution possibly points to the presence of significant photometric heterogeneity in the TDE landscape. AT2019cmw possibly represents a `missing link' between TDEs and the recently categorised LFCs, which \citet{nicholl_at_2023} speculate may be peculiar TDEs from IMBHs or stellar mass black holes.

\end{itemize}

We have found AT2019cmw to be an extreme event in many aspects, with few analogous events in the literature. Our model estimates of a high disrupted star mass, if correct, point to the possibility that TDEs could provide a unique window from which we could probe star formation in close proximity to their host galaxies' SMBH. However, our \textsc{Redback} fits have highlighted the difficulties in measuring SMBH masses by fitting lightcurves using current TDE models, suggesting that additional theoretical work is currently needed to improve their physical accuracy. Many of the driving mechanisms behind aspects of AT2019cmw's and other atypical TDEs' evolution are also currently unknown, such as what causes the presentation of featureless spectra, and why some TDEs appear to cool post-peak whilst others do not. With >35000 new TDE discoveries expected from the Vera C. Rubin observatory \citep{ivezic_lsst_2019,bricman_prospects_2020}, and observatories such as the La Silla Schmidt Southern Survey (LS4; \citealt{miller_silla_2025}) due to begin science operations in the near future, many more AT2019cmw-like events will soon be discovered. Significantly increasing our sample size of well-studied edge cases such as AT2019cmw will allow us to probe the limits of current TDE models. It will also allow us to better study the observational similarities and differences between peculiar and `normal' TDEs, improving our understanding of both sub-populations, as well as enabling us to refine our photometric filtering methods when separating TDEs and supernovae.

\section*{Acknowledgements}

We thank the referee for their helpful suggestions for improving this manuscript.

We thank Brian Metzger, Jean Somalwar, Vikram Ravi and Erica Hammerstein for their assistance in the interpretation of this event.

Blackbody temperatures, radii and luminosities for AT2018jbv, AT2020qhs, AT2020riz and AT2020ysg were supplied by Erica Hammerstein in private communications.

M.W.C acknowledges support from the National Science Foundation with grant numbers PHY-2117997, PHY-2308862 and PHY-2409481.

Based on observations obtained with the Samuel Oschin Telescope 48-inch and the 60-inch Telescope at the Palomar Observatory as part of the Zwicky Transient Facility project. ZTF is supported by the National Science Foundation under Grant No. AST-1440341 and a collaboration including Caltech, IPAC, the Weizmann Institute of Science, the Oskar Klein Center at Stockholm University, the University of Maryland, the University of Washington, Deutsches Elektronen-Synchrotron and Humboldt University, Los Alamos National Laboratories, the TANGO Consortium of Taiwan, the University of Wisconsin at Milwaukee, and Lawrence Berkeley National Laboratories. Operations are conducted by COO, IPAC, and UW.

The ZTF forced-photometry service was funded under the Heising-Simons Foundation grant No. 12540303 (PI: Graham).

Zwicky Transient Facility access was supported by Northwestern University and the Center for Interdisciplinary Exploration and Research in Astrophysics (CIERA).

The Gordon and Betty Moore Foundation, through both the Data-Driven Investigator Program and a dedicated grant, provided critical funding for SkyPortal.

This work has made use of data from the Asteroid Terrestrial-impact Last Alert System (ATLAS) project. The Asteroid Terrestrial-impact Last Alert System (ATLAS) project is primarily funded to search for near earth asteroids through NASA grants NN12AR55G, 80NSSC18K0284, and 80NSSC18K1575; byproducts of the NEO search include images and catalogs from the survey area. This work was partially funded by Kepler/K2 grant J1944/80NSSC19K0112 and HST GO-15889, and STFC grants ST/T000198/1 and ST/S006109/1. The ATLAS science products have been made possible through the contributions of the University of Hawaii Institute for Astronomy, the Queen’s University Belfast, the Space Telescope Science Institute, the South African Astronomical Observatory, and The Millennium Institute of Astrophysics (MAS), Chile.

The Liverpool Telescope is operated on the island of La Palma by Liverpool John Moores University in the Spanish Observatorio del Roque de los Muchachos of the Instituto de Astrofisica de Canarias with financial support from the UK Science and Technology Facilities Council.

The Pan-STARRS1 Surveys (PS1) and the PS1 public science archive have been made possible through contributions by the Institute for Astronomy, the University of Hawaii, the Pan-STARRS Project Office, the Max-Planck Society and its participating institutes, the Max Planck Institute for Astronomy, Heidelberg and the Max Planck Institute for Extraterrestrial Physics, Garching, The Johns Hopkins University, Durham University, the University of Edinburgh, the Queen's University Belfast, the Harvard-Smithsonian Center for Astrophysics, the Las Cumbres Observatory Global Telescope Network Incorporated, the National Central University of Taiwan, the Space Telescope Science Institute, the National Aeronautics and Space Administration under Grant No. NNX08AR22G issued through the Planetary Science Division of the NASA Science Mission Directorate, the National Science Foundation Grant No. AST-1238877, the University of Maryland, Eotvos Lorand University (ELTE), the Los Alamos National Laboratory, and the Gordon and Betty Moore Foundation.

This research has made use of the Keck Observatory Archive (KOA), which is operated by the W. M. Keck Observatory and the NASA Exoplanet Science Institute (NExScI), under contract with the National Aeronautics and Space Administration.

The National Radio Astronomy Observatory is a facility of the National Science Foundation operated under cooperative agreement by Associated Universities, Inc.

This research has made use of the Spanish Virtual Observatory (https://svo.cab.inta-csic.es) project funded by MCIN/AEI/10.13039/501100011033/ through grant PID2020-112949GB-I00

This publication makes use of data products from the Near-Earth Object Wide-field Infrared Survey Explorer (NEOWISE), which is a joint project of the Jet Propulsion Laboratory/California Institute of Technology and the University of Arizona. NEOWISE is funded by the National Aeronautics and Space Administration.

This research has made use of the NASA/IPAC Infrared Science Archive, which is funded by the National Aeronautics and Space Administration and operated by the California Institute of Technology.

%%%%%%%%%%%%%%%%%%%%%%%%%%%%%%%%%%%%%%%%%%%%%%%%%%
\section*{Data Availability}

The data presented in this paper will be provided upon request to the author. Spectra are available to download on \href{https://www.wiserep.org/object/28553}{WISeREP}. Photometry is listed in online supplementary data tables.

%%%%%%%%%%%%%%%%%%%% REFERENCES %%%%%%%%%%%%%%%%%%

% The best way to enter references is to use BibTeX:

\bibliographystyle{mnras}
\bibliography{AT2019cmw} % if your bibtex file is called example.bib

% Alternatively you could enter them by hand, like this:
% This method is tedious and prone to error if you have lots of references
%\begin{thebibliography}{99}
%\bibitem[\protect\citeauthoryear{Author}{2012}]{Author2012}
%Author A.~N., 2013, Journal of Improbable Astronomy, 1, 1
%\bibitem[\protect\citeauthoryear{Others}{2013}]{Others2013}
%Others S., 2012, Journal of Interesting Stuff, 17, 198
%\end{thebibliography}

%%%%%%%%%%%%%%%%%%%%%%%%%%%%%%%%%%%%%%%%%%%%%%%%%%

%%%%%%%%%%%%%%%%% APPENDICES %%%%%%%%%%%%%%%%%%%%%
\FloatBarrier
\appendix

\section{Supplementary data tables}
\label{sec:AppendixC}

Photometry of AT2019cmw, and supplementary LT photometry of AT2018jbv, are presented in Table~\ref{tab:AT2019cmw_phot_table} and Table~\ref{tab:AT2018jbv_phot_table} respectively. Blackbody parameters derived using the methods described in Section~\ref{sec:Blackbody} for AT2019cmw and AT2018jbv are presented in Table~\ref{tab:AT2019cmw_blackbody_table} and Table~\ref{tab:AT2018jbv_blackbody_table}.

\begin{table}
\caption{Optical/UV photometry of AT2019cmw. We list only the first ten entries of AT2019cmw's photometry here, the full data is available in the supplementary online content.}
\label{tab:AT2019cmw_phot_table}
\begin{tabular}{rrrll}
\toprule
         MJD &    mag &  magerr & band & instrument \\
\midrule
58594.143710 & 18.709 &   0.099 &    g &         LT \\
58598.226021 & 18.706 &   0.043 &    g &         LT \\
58603.229402 & 18.731 &   0.048 &    g &         LT \\
58603.101068 & 18.750 &   0.044 &    g &         LT \\
58607.077287 & 18.781 &   0.039 &    g &         LT \\
58608.207288 & 18.833 &   0.038 &    g &         LT \\
58611.058513 & 18.805 &   0.036 &    g &         LT \\
58613.151103 & 18.858 &   0.036 &    g &         LT \\
58615.073996 & 18.930 &   0.039 &    g &         LT \\
58617.212955 & 18.992 &   0.038 &    g &         LT \\
\bottomrule
\end{tabular}
\end{table}

\begin{table}
\caption{Supplementary LT photometry of AT2018jbv. We list only the first ten entries of AT2018jbv's photometry here, the full data is available in the supplementary online content.}
\label{tab:AT2018jbv_phot_table}
\begin{tabular}{rrrll}
\toprule
         MJD &    mag &  mag err & band & telescope \\
\midrule
58547.047175 & 19.332 &    0.073 &    g &        LT \\
58566.029862 & 19.293 &    0.167 &    g &        LT \\
58584.987308 & 19.509 &    0.074 &    g &        LT \\
58595.002000 & 19.484 &    0.124 &    g &        LT \\
58879.239850 & 20.749 &    0.095 &    g &        LT \\
58896.150785 & 20.863 &    0.079 &    g &        LT \\
58925.070166 & 20.932 &    0.080 &    g &        LT \\
58959.973530 & 21.126 &    0.086 &    g &        LT \\
58985.930979 & 21.070 &    0.082 &    g &        LT \\
59024.922616 & 21.092 &    0.091 &    g &        LT \\
\bottomrule
\end{tabular}
\end{table}

\begin{table*}
\centering
\caption{Blackbody characteristics of AT2019cmw derived using the methods described in Section~\ref{sec:Blackbody}.}
\label{tab:AT2019cmw_blackbody_table}
\begin{tabular}{lllllll}
\toprule
Phase &        Temperature &             Temperature Error &         Radius &      Radius error &                Luminosity &              Luminosity error \\
\midrule
days from peak (rest frame) &    $\mathrm{log(T/K)}$ &            $\mathrm{log(T/K)}$ &   $\mathrm{log(R/cm)}$ &   $\mathrm{log(R/cm)}$ &     $\mathrm{log(L/\mathrm{ergs^{-1}})}$ &      $\mathrm{log(L/\mathrm{ergs^{-1}})}$ \\
\midrule
-21.659 & 4.451 & 0.047 & 14.990 & 0.030 & 44.639 & 0.116 \\
-20.342 & 4.451 & 0.047 & 15.130 & 0.020 & 44.918 & 0.100 \\
-15.076 & 4.451 & 0.047 & 15.363 & 0.007 & 45.382 & 0.095 \\
-14.417 & 4.451 & 0.047 & 15.367 & 0.011 & 45.393 & 0.096 \\
-11.784 & 4.451 & 0.047 & 15.413 & 0.006 & 45.483 & 0.095 \\
 -3.884 & 4.451 & 0.047 & 15.471 & 0.003 & 45.599 & 0.094 \\
 -2.567 & 4.451 & 0.047 & 15.473 & 0.006 & 45.605 & 0.095 \\
 -1.909 & 4.451 & 0.047 & 15.475 & 0.002 & 45.608 & 0.094 \\
  4.674 & 4.441 & 0.039 & 15.466 & 0.036 & 45.550 & 0.171 \\
  7.966 & 4.463 & 0.045 & 15.434 & 0.040 & 45.572 & 0.198 \\
 10.599 & 4.478 & 0.020 & 15.413 & 0.021 & 45.590 & 0.091 \\
 13.232 & 4.514 & 0.054 & 15.375 & 0.046 & 45.659 & 0.234 \\
 14.549 & 4.284 & 0.010 & 15.594 & 0.016 & 45.175 & 0.050 \\
 15.866 & 4.439 & 0.038 & 15.427 & 0.036 & 45.461 & 0.169 \\
 20.474 & 4.251 & 0.037 & 15.646 & 0.091 & 45.151 & 0.233 \\
 22.449 & 4.300 & 0.055 & 15.547 & 0.061 & 45.148 & 0.254 \\
 23.766 & 4.277 & 0.009 & 15.561 & 0.013 & 45.082 & 0.046 \\
 26.399 & 4.354 & 0.019 & 15.462 & 0.024 & 45.192 & 0.091 \\
 44.832 & 4.166 & 0.044 & 15.573 & 0.058 & 44.661 & 0.212 \\
 52.074 & 4.171 & 0.027 & 15.561 & 0.036 & 44.659 & 0.130 \\
 63.265 & 4.182 & 0.013 & 15.501 & 0.021 & 44.584 & 0.067 \\
 98.157 & 4.052 & 0.029 & 15.634 & 0.050 & 44.332 & 0.153 \\
115.932 & 4.131 & 0.065 & 15.502 & 0.100 & 44.382 & 0.326 \\
146.215 & 4.047 & 0.023 & 15.570 & 0.040 & 44.179 & 0.121 \\
203.489 & 3.930 & 0.045 & 15.704 & 0.093 & 43.982 & 0.260 \\
210.072 & 3.986 & 0.029 & 15.564 & 0.055 & 43.927 & 0.159 \\
229.164 & 4.017 & 0.041 & 15.447 & 0.082 & 43.814 & 0.232 \\
242.989 & 3.891 & 0.054 & 15.707 & 0.118 & 43.832 & 0.319 \\
296.313 & 3.836 & 0.057 & 15.704 & 0.130 & 43.604 & 0.345 \\
\bottomrule
\end{tabular}
\end{table*}

\begin{table*}
\centering
\caption{Blackbody characteristics of AT2018jbv derived using the methods described in Section~\ref{sec:Blackbody}. The phase of each fitted epoch is set according to the peak time of MJD 58469.19 for AT2018jbv from \protect\citet{hammerstein_final_2023}.}
\label{tab:AT2018jbv_blackbody_table}
\begin{tabular}{lllllll}
\toprule
  Phase &  Temperature &  Temperature Error &  Radius &  Radius Error &  Luminosity &  Luminosity Error \\
\midrule
days from peak (rest frame) &    $\mathrm{log(T/K)}$ &            $\mathrm{log(T/K)}$ &   $\mathrm{log(R/cm)}$ &   $\mathrm{log(R/cm)}$ &     $\mathrm{log(L/\mathrm{ergs^{-1}})}$ &      $\mathrm{log(L/\mathrm{ergs^{-1}})}$ \\
\midrule
 53.590 &        4.463 &              0.028 &  15.210 &         0.041 &      45.123 &             0.141 \\
 58.067 &        4.452 &              0.083 &  15.158 &         0.071 &      44.977 &             0.359 \\
 68.515 &        4.425 &              0.029 &  15.229 &         0.043 &      45.012 &             0.145 \\
 72.246 &        4.274 &              0.133 &  15.346 &         0.130 &      44.640 &             0.593 \\
 78.963 &        4.398 &              0.037 &  15.228 &         0.060 &      44.902 &             0.191 \\
 84.187 &        4.397 &              0.037 &  15.232 &         0.060 &      44.905 &             0.191 \\
 85.679 &        4.414 &              0.102 &  15.152 &         0.089 &      44.811 &             0.444 \\
 89.410 &        4.338 &              0.031 &  15.333 &         0.053 &      44.872 &             0.163 \\
 93.888 &        4.424 &              0.138 &  15.152 &         0.112 &      44.854 &             0.596 \\
 94.634 &        4.596 &              0.104 &  14.972 &         0.087 &      45.180 &             0.450 \\
114.784 &        4.219 &              0.303 &  15.336 &         0.194 &      44.400 &             1.273 \\
158.813 &        4.325 &              0.039 &  15.214 &         0.067 &      44.581 &             0.207 \\
305.828 &        4.557 &              0.207 &  14.785 &         0.129 &      44.652 &             0.866 \\
318.515 &        4.465 &              0.132 &  14.844 &         0.103 &      44.399 &             0.567 \\
340.157 &        4.347 &              0.100 &  14.944 &         0.092 &      44.127 &             0.439 \\
365.530 &        4.333 &              0.124 &  14.919 &         0.116 &      44.022 &             0.548 \\
384.933 &        4.498 &              0.263 &  14.788 &         0.158 &      44.423 &             1.097 \\
414.037 &        4.540 &              0.219 &  14.729 &         0.129 &      44.469 &             0.912 \\
\bottomrule
\end{tabular}
\end{table*}

\section{Cooling envelope model posterior distributions}
\label{sec:AppendixA}

\begin{figure*}
\begin{center}
    \includegraphics[width=2\columnwidth]{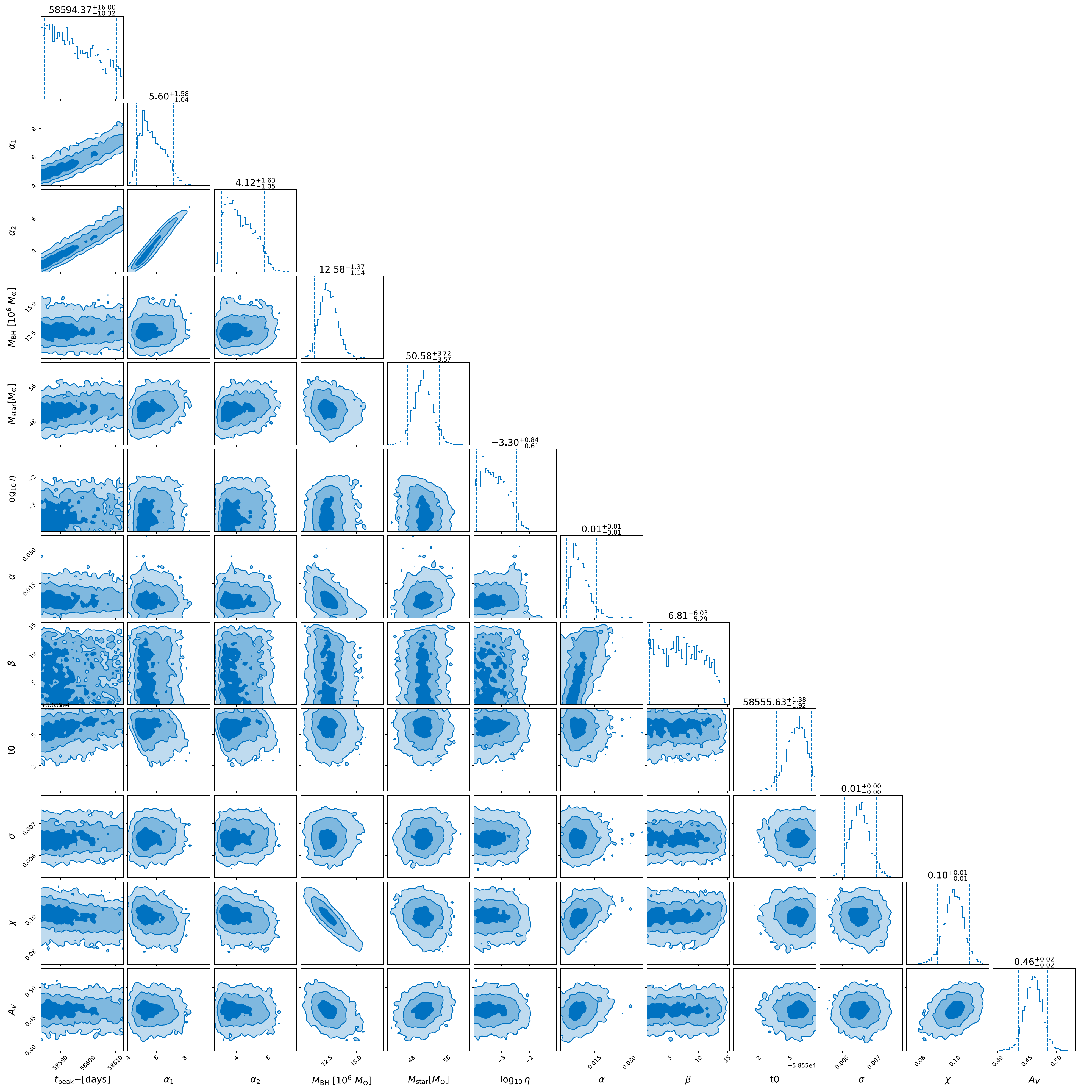}
\end{center}
\caption{\label{fig:CornerMultibandFull} Inferred parameters from our model fit detailed in Section~\ref{sec:Redback}. Plot made using \textsc{corner} \citep{foreman-mackey_cornerpy_2016}.}
\end{figure*}

\begin{figure*}
\begin{center}
    \includegraphics[width=2\columnwidth]{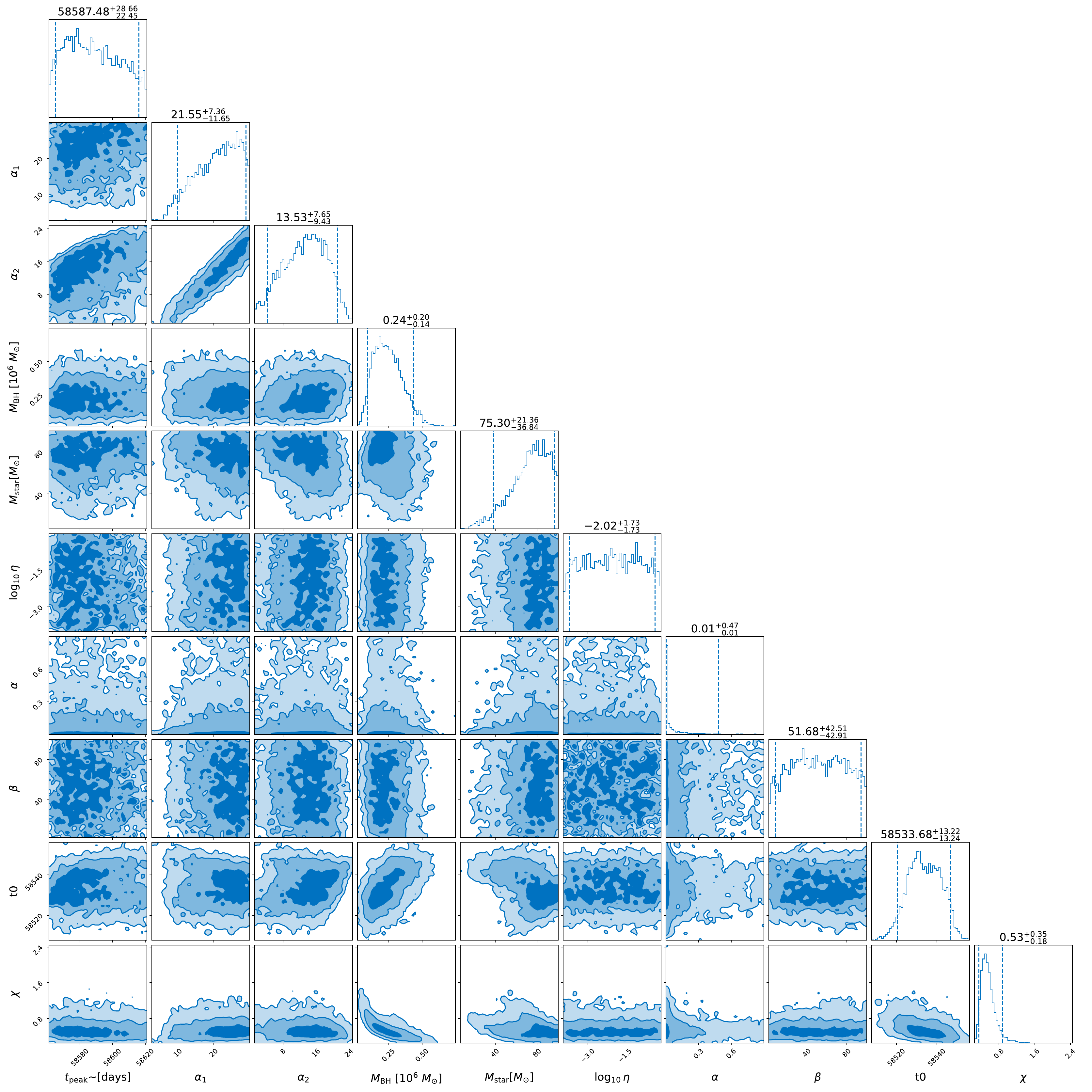}
\end{center}
\caption{\label{fig:CornerBolometricFull} Inferred parameters from our model fit detailed in Section~\ref{sec:RedbackBolometric}. Plot made using \textsc{corner} \citep{foreman-mackey_cornerpy_2016}.}
\end{figure*}

Figure~\ref{fig:CornerMultibandFull} shows the full posterior distribution inferred by our cooling envelope model fit to AT2019cmw's multiband photometry, as detailed in Section~\ref{sec:Redback}. Figure~\ref{fig:CornerBolometricFull} also shows the full posterior distribution inferred by our fit to our derived bolometric luminosities for AT2019cmw from Section~\ref{sec:Blackbody}, as detailed in Section~\ref{sec:RedbackBolometric}.

\section{Cooling envelope model predicted characteristics}
\label{sec:AppendixB}

\begin{figure}
\begin{center}
    \includegraphics[width=1\columnwidth]{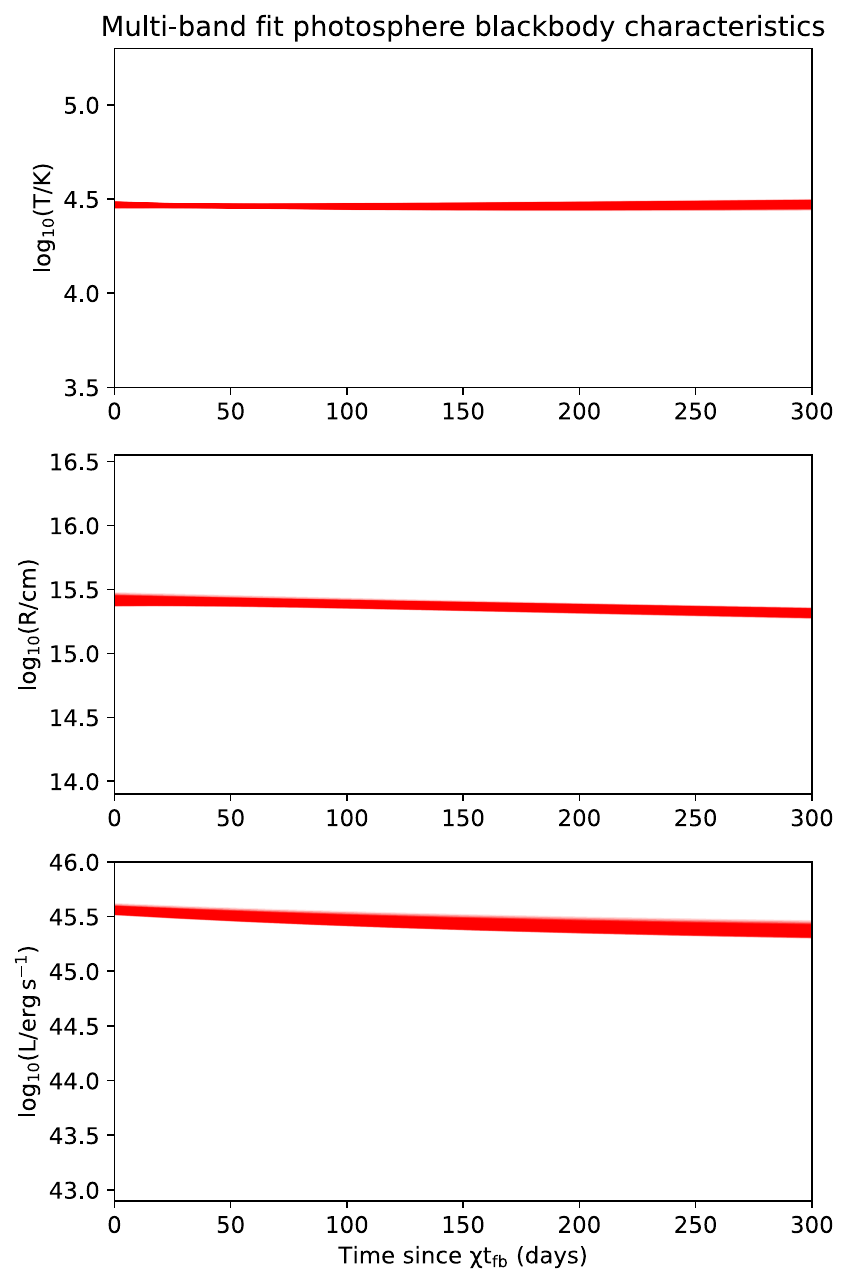}
\end{center}
\caption{\label{fig:MultibandBlackbody} Photospheric blackbody temperatures (\textbf{top}), radii (\textbf{middle}) and luminosities (\textbf{bottom}) predicted for AT2019cmw by our model fit from Section~\ref{sec:Redback} up to 300 days post-envelope formation.}
\end{figure}

\begin{figure}
\begin{center}
    \includegraphics[width=1\columnwidth]{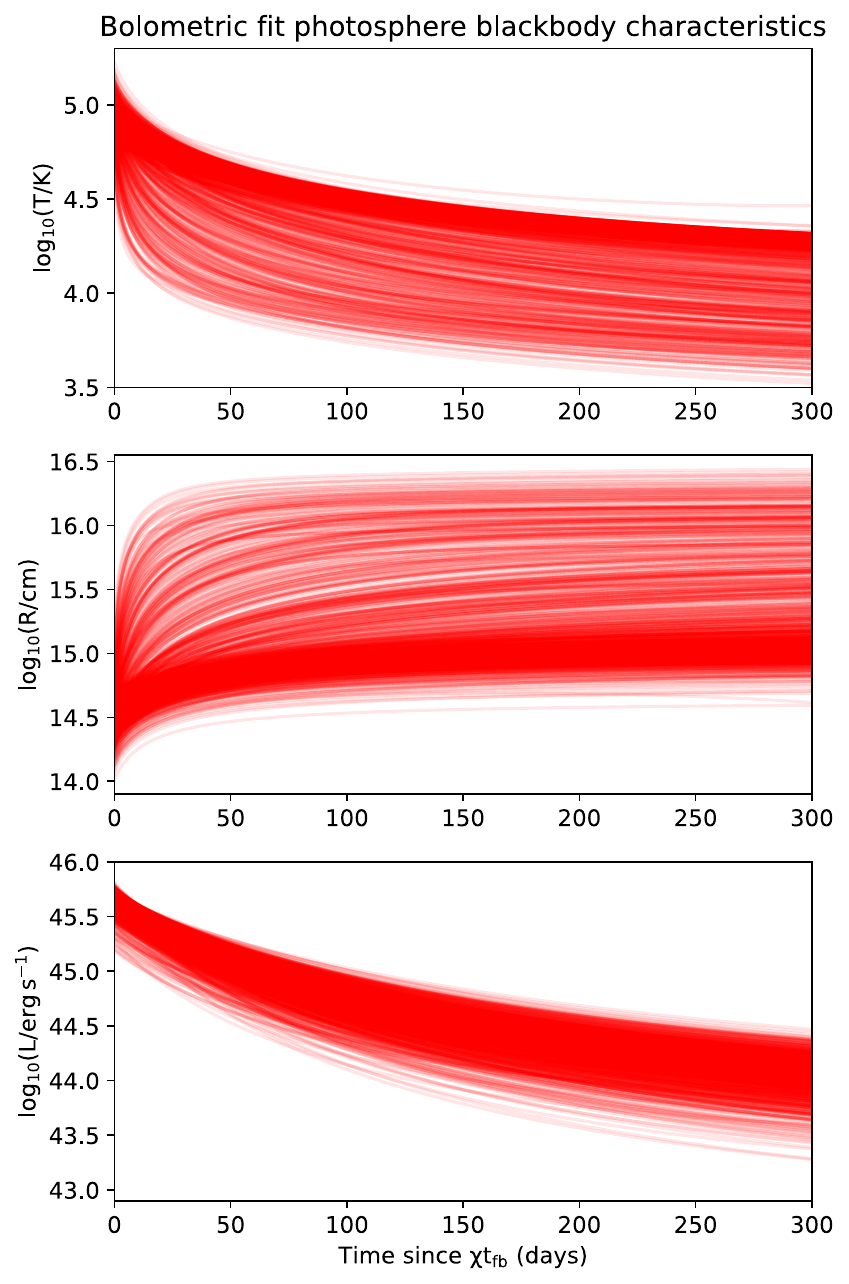}
\end{center}
\caption{\label{fig:BolometricBlackbody} Photospheric blackbody temperatures (\textbf{top}), radii (\textbf{middle}) and luminosities (\textbf{bottom}) predicted for AT2019cmw by our model fit from Section~\ref{sec:RedbackBolometric} up to 300 days post-envelope formation.}
\end{figure}

\begin{figure}
\begin{center}
    \includegraphics[width=1\columnwidth]{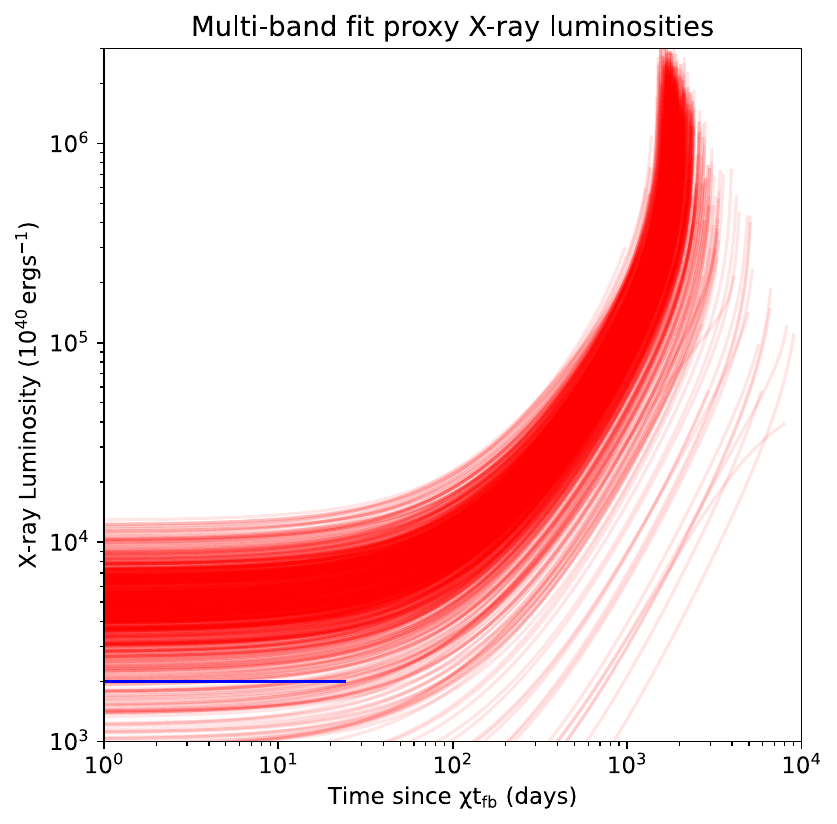}
\end{center}
\caption{\label{fig:ProxyXray} Proxy X-ray luminosities predicted for AT2019cmw by our model fit from Section~\ref{sec:Redback} up to $10^4$ days post-envelope formation, shown in red. Our upper limit on AT2019cmw's X-ray luminosity from Section~\ref{sec:XrayResults} is shown as a blue horizontal line.}
\end{figure}

Figure~\ref{fig:MultibandBlackbody} and Figure~\ref{fig:BolometricBlackbody} show the photospheric blackbody characteristics predicted by our cooling envelope model fits from Section~\ref{sec:Redback} and Section~\ref{sec:RedbackBolometric} respectively. We also present proxy X-ray luminosities predicted by our model fit from Section~\ref{sec:Redback} in Figure~\ref{fig:ProxyXray}.

%%%%%%%%%%%%%%%%%%%%%%%%%%%%%%%%%%%%%%%%%%%%%%%%%%

% Don't change these lines
\bsp	% typesetting comment
\label{lastpage}
\end{document}